\documentclass[a4paper,12pt,english]{article}

\usepackage{amsfonts,bm,amssymb,euscript,array,babel,cite}
\usepackage{mathtools}
\usepackage{tikz-cd}
\usepackage{amsmath}
\usepackage{tensor}
\usepackage{hhline}
\usepackage[amsmath]{empheq}
\usepackage{hyperref}
\usepackage{cancel}
\usepackage{mathrsfs}
\usepackage{pdfsync}
\usepackage{multirow}
\usepackage{tikz}

\usepackage{framed}

\newsavebox{\mysaveboxM}
\newsavebox{\mysaveboxT}

\makeatletter

\newcommand{\dd}{\mathrm{d}}

\newcommand{\w}{\wedge}

\newcommand{\be}{\begin{equation}}
\newcommand{\ee}{\end{equation}}

\def\nn{\nonumber}
\def \bea{\begin{eqnarray}} 
\def\eea{\end{eqnarray}}

\def\bi{\begin{itemize}} 
\def\ei{\end{itemize}}

\def\bs{\boldsymbol}

\makeatother

\def\a{\alpha} \def\b{\beta}  \def\G{\Gamma} \def\d{\delta} \def\D{\Delta}
\def\e{\epsilon} 
   
  \def\m{\mu}
    
\def\s{\sigma} \def\S{\Sigma}

\def\one{\mbox{1 \kern-.59em {\rm l}}}

\numberwithin{equation}{section}

\begin{document}

\makeatother
\parindent=0cm
\renewcommand{\title}[1]{\vspace{10mm}\noindent{\Large{\bf #1}}\vspace{8mm}} \newcommand{\authors}[1]{\noindent{\large #1}\vspace{5mm}} \newcommand{\address}[1]{{\itshape #1\vspace{2mm}}}

\begin{titlepage}

\begin{flushright}
	RBI-ThPhys-2022-22
\end{flushright}

\begin{center}

\title{ {\Large {The BV action of 3D twisted R-Poisson sigma models}}}

  \authors{\large Athanasios {Chatzistavrakidis\,}$^{\a}$, Noriaki Ikeda\,$^{\b}$, Grgur \v{S}imuni\'c\,$^{\a}$ 
  }
 
 
  \address{ $^{\a}$ Division of Theoretical Physics, Rudjer Bo\v skovi\'c Institute \\ Bijeni\v cka 54, 10000 Zagreb, Croatia \\
 
 \
 
$^{\b}$Department of Mathematical Sciences, Ritsumeikan University \\
Kusatsu, Shiga 525-8577, Japan
 
}

\vskip 2cm

\begin{abstract}
We determine the solution to the classical master equation for a 3D topological field theory with Wess-Zumino term and an underlying geometrical structure of a twisted R-Poisson manifold on its target space. The graded geometry of the target space departs from the usual QP structure encountered in the AKSZ construction of topological sigma models, the obstruction being attributed to the presence of the Wess-Zumino 4-form. Due to the inapplicability of the AKSZ construction in this case, we set up the traditional BV/BRST formalism for twisted R-Poisson sigma models in any dimension, which feature an open gauge algebra and constitute multiple stages reducible constrained Hamiltonian systems. An unusual feature of the theories is that it exhibits non-linear openness of the gauge algebra, in other words products of the equations of motion appear in it. Nevertheless, we find the BV action in presence of the 4-form twist in 3D, namely for a specific 4-form twisted (pre-)Courant sigma model. Moreover, we provide a complete set of explicit formulas for the off-shell nilpotent BV operator for untwisted R-Poisson sigma models in any dimension.  
\end{abstract}

\end{center}

\vskip 2cm

\end{titlepage}

\setcounter{footnote}{0}

\newpage

\tableofcontents

\section{Introduction}
\label{sec1}

The BV/BRST formalism was developed as a general method to quantize an arbitrary field theory with gauge redundancies \cite{Batalin:1981jr,Batalin:1983ggl}.{\footnote{Useful expositions of the subject include the book \cite{HT} and the review \cite{Gomis:1994he}.}} It is particularly useful, and in fact necessary, for field theories whose gauge structure is reducible, namely when not all gauge transformations are independent, or whose gauge algebra closes either only up to equations of motion or with structure ``constants'' that depend on the fields of the theory. This is the case in many supersymmetric field theories or supergravity, string field theory, more generally any time when differential form fields of degree higher than one are present in the theory, but also in a variety of topological field theories such as the A-model, {B-model} \cite{Witten:1991zz,Witten:1988xj}, Chern-Simons theory \cite{Deser:1981wh,Witten:1988hf} {and BF theories}.

The construction of topological quantum field theory in the spirit of the BV/BRST formalism found a geometric foundation through the work of Alexandrov, Kontsevich, Schwarz and Zaboronsky, who showed that solutions to the classical master equation correspond to graded supermanifolds with a QP structure \cite{Alexandrov:1995kv}. QP manifolds are characterised by the existence of a cohomological vector field Q of degree 1 and of a graded symplectic structure P, which are compatible in the sense that the graded symplectic form is Q-invariant. The AKSZ construction utilizes this structure to build a ladder of topological field theories that correspond to the geometric ladder of QP manifolds in diverse dimensions. The first instances include the 2D Poisson sigma model \cite{SchallerStrobl,Ikeda}---and the A-model {and B-model} through its gauge fixing---and an extension of the 3D Chern-Simons theory from algebras to algebroids, called the Courant sigma model \cite{Ikeda:2000yq,Ikeda:2002wh,Hofman:2002jz,Roytenberg:2006qz}. Both have found numerous applications, notably in the deformation quantization of Poisson manifolds \cite{Kontsevich:1997vb,Cattaneo:1999fm} which to some extend sparked the development of noncommutative field theory \cite{Szabo:2001kg} and in the description of general string backgrounds with fluxes \cite{Mylonas:2012pg,Chatzistavrakidis:2015vka,Chatzistavrakidis:2018ztm,Bessho:2015tkk,Heller:2016abk} and T-duality \cite{Severa:2016prq}. 

On the other hand, it is fair to say that the BV/BRST formalism is more general than the AKSZ construction, since it is possible to imagine that the target space of a given gauge field theory might not possess an underlying QP structure. This is indeed the case in a variety of topological field theories, such as the 2D H-twisted WZW-Poisson sigma model \cite{Klimcik:2001vg}, the 2D Dirac sigma models \cite{Kotov:2004wz} and a class of models with twisted R-Poisson structure in any spacetime dimension that was recently described in Ref. \cite{Chatzistavrakidis:2021nom} and extended further in Ref. \cite{Ikeda:2021rir} to higher Dirac structures. The underlying reason for the obstruction to QP-ness can be either the presence of a Wess-Zumino term that renders the Q and P structures of the graded manifold incompatible or, more drastically, the very absence of a P structure.{\footnote{See for example \cite{Alkalaev:2013hta} for the case of presymplectic structures.}} In such cases the original AKSZ construction does not apply. 

Nevertheless, it was demonstrated recently in the case of the H-twisted Poisson sigma model that a solution to the classical master equation can be determined  \cite{Ikeda:2019czt}. This solution, found using the traditional approach to the BV formalism, involves a single term quadratic in antifields, which modifies the one of the untwisted case by $H$-dependent terms, $H$ being the closed 3-form on the target space that gives rise to the Wess-Zumino term. The full term is controlled by the so-called basic curvature associated to the notion of an $E$-connection in the case $E=T^{\ast}M$. $E$-connections and $E$-covariant derivatives generalize the usual notion of vector bundle connections and covariant derivatives in case the tangent bundle $TM$ is replaced by a general vector bundle $E$ with suitable structure, for example a Lie or Courant algebroid {\cite{Blaom, Abad-Crainic, Kotov-Strobl2}}.
Such higher geometric structures appear to play a central role in the BV/BRST formalism, notably also for topological field theories without a QP structure in their target space.     

The main goal of the present paper is to determine the solution to the classical master equation for 4-form-twisted R-Poisson sigma models in 3D. 
This is a special case of a general class of topological field theories with a $(p+2)$-form-twisted R-Poisson structure on the target space in any world volume dimension $p+1$. This structure comprises a smooth target space manifold $M$ endowed with a Poisson bivector $\Pi\in \G(\w^2TM)$ together with a fully antisymmetric multivector $R$ of order $p+1$ and a closed $(p+2)$-form $H$ such that{\footnote{We have corrected a minus sign in comparison to the definition in Ref. \cite{Chatzistavrakidis:2021nom}.}} 
\be 
[\Pi,R]=(-1)^{p+1}\langle\otimes^{p+2}\Pi, H\rangle, 
\ee 
where the bracket on the left-hand side is the Schouten-Nijenhuis bracket of multivector fields. Thus a twisted R-Poisson manifold is given as the quadruple $(M,\Pi,R,H)$. For vanishing $R$ and $H$ it reduces to a Poisson manifold and thus higher dimensional generalizations of the Poisson sigma model are included as special cases. 

A graded-geometric incarnation of the structure is achieved by considering the graded supermanifold ${\cal M}=T^{\ast}[p]T^{\ast}[1]M$ equipped with a cohomological vector field, a Q-structure. Being a degree-shifted cotangent bundle, i.e. a generalized phase space, this manifold has a natural graded symplectic structure P. When $H$ vanishes, the two structures are compatible, rendering ${\cal M}$ a QP manifold, however this ceases to be true in presence of the $(p+2)$-form $H$, to which we refer as the twist. 

Once a sigma model with this target space is written down, the twist appears as a Wess-Zumino term. The theory contains four different types of fields, scalars $X^{i}$, spacetime 1-forms $A_i$, spacetime $(p-1)$-forms $Y^{i}$ and spacetime $p$-forms $Z_i$, the index being one on the target space. Since there are differential forms of degree higher than 1, the theory is multiple stages reducible and moreover its gauge algebra does not close off-shell. This means that the BV formalism must be used to quantize the theory. As already mentioned, this is a standardized task in absence of Wess-Zumino term, since one can simply apply the ASKZ construction. For instance, the theory in 3D is a specific version of the Courant sigma model. 

On the other hand, turning on the Wess-Zumino term leads to a departure from this simple picture. Since there is no standardized procedure \`a la AKSZ for determining the BV action in such cases, one should find it in a direct way, as in the case of the $H$-twisted Poisson sigma model. However, comparing to this 2D case, new features appear in higher dimensions. Notably, we will see that the square of the BRST operator on the top form $Z_i$ contains field equations not only linearly but also non-linearly, namely products of field equations. This is not the case in 2D models, like the $H$-twisted Poisson sigma model. Nevertheless, a careful application of the BV formalism leads to the correct BV action in the 3D case, with all modifications due to the twist taken into account. This is then the first example of 4-form twisted Courant sigma model, in the sense of \cite{Hansen:2009zd}, whose BV action is determined. The mathematical structure underlying this model is within the class of pre-Courant algebroids \cite{Vaisman:2004msa,preca2,prehequiv}. 

The problem can be actually formulated in arbitrary world volume dimensions. Using the gauge invariances of the classical action, the ghost and ghost-for-ghost content of the theory is easily identified and the BRST operator and its square can be calculated for the full tower of fields and ghosts in presence of the $(p+2)$-form twist. As one would expect, the square of the BRST operator turns out to be nonvanishing yet proportional to the classical field equations or products thereof. A genuine BV operator, i.e. one that is strictly nilpotent, then requires the introduction of antifields in the theory. In general dimensions, we provide a universal set of closed formulas for the BV operator on all fields and ghosts in the case of vanishing twist. Although such expressions have been written down explicitly for 3D Courant sigma models \cite{Ikeda:2000yq} (see also \cite{Grewcoe:2020ren} for an $L_{\infty}$ approach), here we derive an elegant set of formulas for topological sigma models in any dimension. The essential content of these formulas could also be found via the AKSZ construction in the untwisted case, nevertheless we have determined them in a direct way, which results in very explicit and compact expressions. As already mentioned, the problem becomes more tractable for 3 world volume dimensions, in which case we determine the full BV operator including the twist and consequently the full BV action for 4-form-twisted Courant sigma models of twisted R-Poisson type.     

The rest of the paper is organised as follows. In Section \ref{sec2} we describe the main features of twisted R-Poisson sigma models and the geometry of their target space. In Section \ref{sec21} twisted R-Poisson manifolds are defined and their description as Q-manifolds in graded geometry is explained. The corresponding topological fields theory is discussed in Section \ref{sec22} together with its gauge symmetries and field equations. Section \ref{sec31} contains the identification of the ghosts and ghosts for ghosts that must be included in the theory in arbitrary dimensions. In the same section we determine the BRST operator for all fields and ghosts and show that it is nilpotent only on-shell. We proceed in Section \ref{sec32} with the introduction of antifields and antighosts and the extension of the BRST operator to the BV operator, which is nilpotent off-shell, in the untwisted case. We then focus on 3-dimensional world volumes with 4-form Wess-Zumino term in Section \ref{sec4}, where we fully determine all the $H$-dependent corrections to the AKSZ R-Poisson-Courant sigma model. Finally, Section \ref{sec5} contains our conclusions and outlook.

\section{Twisted R-Poisson topological field theory}
\label{sec2}

\subsection{Twisted R-Poisson manifolds}
\label{sec21}

The geometrical structure of the target space for the topological field theories considered in this paper is a twisted R-Poisson manifold. This is an extension of Poisson and twisted Poisson structures to include multisymplectic structures. Before explaining it in more detail, let us briefly recall some basic facts about ordinary Poisson geometry that will be useful in the ensuing. Our goal is to characterize a Poisson structure in several different yet equivalent ways. The most common one is the Lie bracket characterization, where a Poisson manifold is a smooth manifold $M$ equipped with a bilinear, skew symmetric map $\{\cdot,\cdot\}: C^{\infty}(M)\times C^{\infty}(M)\to C^{\infty}(M)$ that satisfies the Jacobi identity and the Leibniz rule. In our approach we will mostly work with the equivalent definition of a Poisson manifold $(M,\Pi)$ with $\Pi\in\G(\wedge^2 TM)$ an antisymmetric bivector field (Poisson structure) that satisfies 
\be 
[\Pi,\Pi]=0
\ee  
with respect to the Schouten-Nijenhuis bracket on multivector fields $[\cdot,\cdot]:\wedge^pTM\times \wedge^q TM\to \wedge^{p+q-1}TM$. It is worth mentioning that this structure can be twisted in a specific way once a closed 3-form $H_3$ on $M$ is considered. This gives rise to the so-called twisted Poisson manifold $(M,\Pi,H_3)$ \cite{Severa:2001qm} with defining conditions 
\bea 
\label{twistedPoisson}
\frac 12 \, [\Pi,\Pi]&=&\langle \otimes^{3}\Pi,H_3\rangle\,, \\[4pt] \dd H_3&=&0\,,
\eea 
where $\dd$ is the de Rham differential and the contraction of the tensor field  $\otimes^3\Pi$ with the 3-form in a given local coordinate system is taken in the odd order indices of each $\Pi$ .

Both above structures may be seen as the geometry underlying certain Lie algebroids. Recall that a Lie algebroid is a vector bundle $E$ over $M$ with a Lie algebra structure on its sections, so that there exists a Lie bracket $[\cdot,\cdot]_{E}:E\times E\to E$ that satisfies a Leibniz rule, like the ordinary Lie bracket of vector fields on the tangent bundle $TM$, with the help of a smooth bundle (anchor) map $\rho: E\to TM$:
\be 
[e,f e']_E=f[e,e']_E+\rho(e)f \, e'\,, \quad e,e'\in \G(E), f\in C^{\infty}(M)\,.
\ee 
Poisson structures and their twisted extension are related to this concept when one makes the choice that the vector bundle is the cotangent bundle, namely $E=T^{\ast}M$, the map $\rho$ is the ``musical'' isomorphism $\Pi^{\sharp}:T^{\ast}M\to TM$ induced by a (possibly twisted, in the above sense) Poisson structure on $M$ and the bracket on sections of the cotangent bundle (that is, 1-forms) is the (twisted) Koszul bracket{\footnote{Note that the action of $\Pi^{\sharp}$ on $e$ is $\Pi^{\sharp}(e)=\Pi^{ij}e_{i}\partial_j.$}} 
\be 
[e,e']_K={\cal L}_{\Pi^{\sharp}(e)}e'-{\cal L}_{\Pi^{\sharp}(e')}e-\dd (\Pi(e,e'))-H_3(\Pi^{\sharp}(e),\Pi^{\sharp}(e'))\,.
\ee 
Then $(T^{\ast}M,\Pi^{\sharp},[\cdot,\cdot]_{K})$ is a Lie algebroid if and only if the defining conditions \eqref{twistedPoisson} hold. 

Alternatively, a more modern way to think about the above structures is in the context of graded differential geometry and $Q$-structures. Specifically, Vaintrob showed in Ref. \cite{Vaintrob} that a Lie algebroid on $E$ is in one-to-one correspondence with a $Q$-manifold ${\cal M}=E[1]$, that is a graded manifold whose fiber coordinates are assigned degree 1, equipped with a cohomological vector field $Q_{E}$ of degree 1, namely one that satisfies $Q^{2}=\frac 12\{Q,Q\}=0$. In the present case, the graded manifold is $T^{\ast}[1]M$ and the degree 1 vector field reads 
\be \label{Qhp}
Q_{T^{\ast}M}=\Pi^{ij}(x)\xi_i\partial_{x^{j}}-\frac 12 (\partial_i\Pi^{jk}+\Pi^{jl}\Pi^{km}H_{ilm})\xi_{j}\xi_{k}\partial_{\xi_{i}}\,,
\ee     
where $(x^{i},\xi_{i})$ are degree 0 and 1 coordinates on the graded manifold respectively and we introduced the notation $\partial_{x^{i}}=\partial/\partial x^{i}$ and $\partial_{\xi_{i}}=\partial/\partial\xi_{i}$. In accordance with the above statements, this vector field is of degree 1 and it satisfies the condition $Q^{2}=0$ if and only if \eqref{twistedPoisson} hold, with or without $H_3$. 

After this very brief introduction to twisted Poisson manifolds and their corresponding Lie algebroids, let us move on to the main concept underlying the field theories we consider in this paper, namely twisted R-Poisson manifolds. These are equipped with the additional structure of a fully antisymmetric multivector field $R$ of order $p+1$. This can give rise to a bracket that generalizes the Poisson bracket{\footnote{Note that every multivector field defines a multiderivation on a manifold, namely a multilinear map $C^{\infty}(M)\times\dots C^{\infty}(M)\to C^{\infty}(M)$ which is totally antisymmetric and a $C^{\infty}$-derivation in each of the arguments \cite{Pham}. In addition, the space of multiderivations and the space of multivector fields of the same order are in one-to-one correspondence.}} $\{\cdot,\cdot\}$, however we directly describe the structure in terms of the alternative formulation based on the Schouten-Nijenhuis bracket. Therefore, we consider the quadruple $(M,\Pi,R,H)$ consisting of a smooth manifold $M$ equipped with a bivector $\Pi\in\G(\wedge^{2}TM)$, an antisymmetric multivector $R\in \G(\wedge^{p+1}TM)$ of degree $p+1$  
and a $(p+2)$-form $H\in\G(\wedge^{p+2}T^{\ast}M)$. This is called a twisted R-Poisson manifold of order $p+1$ when the following conditions hold \cite{Chatzistavrakidis:2021nom}
\bea 
[\Pi,\Pi]&=&0\,, \\[4pt] [\Pi,R]&=&(-1)^{p+1}\langle \otimes^{p+2}\Pi,H\rangle\,, \\[4pt]  \dd H&=&0\,.
\eea 
For completeness and in absence of any other structure one may include in the definition the requirement that $[R,R]=0$ with respect to the Schouten-Nijenhuis bracket, although this does not appear in the field theoretic incarnation of twisted R-Poisson target spaces. Evidently, for vanishing $H$ this should be called an (ordinary or untwisted) R-Poisson structure. Moreover, when in addition $R$ is absent, this reduces to a Poisson structure.{\footnote{We note that an extension of the above to what is called a bi-twisted R-Poisson structure is possible in special cases and reduces to a \emph{twisted} Poisson structure in absence of $H$ and $R$. We refer to \cite{Chatzistavrakidis:2021nom} for more details, since we are not dealing with this more general situation in the present paper.}}  

As already discussed in \cite{Chatzistavrakidis:2021nom}, a twisted R-Poisson structure has a characterization in terms of $Q$-manifolds too. Instead of the degree-shifted cotangent bundle that was associated to the Poisson case, one should now consider the degree-shifted second-order bundle 
\be 
{\cal M}=T^{\ast}[p]T^{\ast}[1]M\,.
\ee 
In a local patch, this manifold can be described by four types of graded coordinates $(x^{i},a_{i},y^{i},z_{i})$ of degrees $(0,1,p-1,p)$ respectively. 
The $Q$-structure on this graded manifold is given by the degree 1 vector field
	\bea 
	Q&=&\Pi^{ji}a_j\partial_{x^{i}}-\frac 12 \partial_i\Pi^{jk}a_ja_k\partial_{a_{i}}+\left((-1)^{p}\Pi^{ji}z_j-\partial_j\Pi^{ik}a_ky^{j}+\frac 1{p!}R^{ij_1\dots j_p}a_{j_1}\dots a_{j_p}\right)\partial_{y^{i}} \, + \nn\\[4pt] 
	&& +\, \left(\partial_i\Pi^{jk}a_kz_j-\frac {(-1)^{p}}2 \partial_i\partial_j\Pi^{kl}y^{j}a_ka_l+\frac {(-1)^{p}}{(p+1)!} f_{i}^{k_1\dots k_{p+1}}a_{k_1}\dots a_{k_{p+1}}\right)\partial_{z_{i}}\,,\label{Qp+1}
	\eea 
	where $f_{i}^{k_1\dots k_{p+1}}= \partial_iR^{k_1\dots k_{p+1}}+\prod_{r=1}^{p+1}\Pi^{k_rl_r}H_{il_1\dots l_{p+1}}$. This vector field is cohomological if and only if $(M,\Pi,R,H)$ is a twisted R-Poisson manifold of order $p+1$ \cite{Chatzistavrakidis:2021nom}.

In both cases above, namely the twisted Poisson and twisted R-Poisson structures, the graded $Q$-manifold we described possesses a graded symplectic structure P given in terms of a graded symplectic 2-form $\omega$. This is evident from the fact that the underlying graded manifold is a cotangent bundle and therefore the graded coordinates form pairs of generalized ``coordinates'' and ``momenta'' as in ordinary Hamiltonian mechanics; such pairs are the $(x,\xi)$ in the (twisted) Poisson case and the $(x,z)$ and $(a,y)$ in the (twisted) R-Poisson case. The graded symplectic structure is of degree 2 in the first case and of degree $p$ in the second case. One may then ask whether the graded manifold has a QP structure. As mentioned in the introduction, this is true if and only if the twist $H$ vanishes. Then the graded symplectic 2-form is indeed $Q$-invariant, namely its Lie derivative along the vector field $Q$ vanishes. However, the presence of $H$ introduces an obstruction to this invariance, as explained in detail in \cite{Ikeda:2019czt} and \cite{Chatzistavrakidis:2021nom} respectively for each of the two cases, and a genuine QP structure does not exist.

\subsection{Twisted R-Poisson sigma models}
\label{sec22}

Given a twisted R-Poisson structure of order $p+1$, there exists a topological field theory in $p+1$ dimensions with target space the corresponding twisted R-Poisson manifold $M$ \cite{Chatzistavrakidis:2021nom}. The fields of the theory are of four different types, specifically (a) a set of scalar fields $X^{i}, i=1,\dots \text{dim}\,M$, which are identified with the components of a sigma model map $X: \S_{p+1} \to M$, where $\S_{p+1}$ is the $(p+1)$-dimensional spacetime where the theory is defined (the world volume; in the few instances when we consider a local coordinate system on it, we refer to its coordinates as $\s^{\m}$ with $\m=0,\dots p$), (b) world volume 1-forms $A_{i}=A_{i\m}(\s)\dd \s^{\m}$ taking values in the pull-back bundle  $X^{\ast}T^{\ast}M$, (c) world volume $(p-1)$-forms $Y^{i}$ taking values in the pull-back bundle $TM$ and (d) world volume $p$-forms $Z_i$ taking values in the pull-back bundle $T^{\ast}M$. Summarizing, the field content of the theory is 
\be 
(X^{i},A_i,Y^{i},Z_i) \quad \text{of form degrees} \quad (0,1,p-1,p)\,.
\ee    

With the above field content, one can write down a general action functional in $p+1$ dimensions with $p\ge 1$, which has the form of a topological sigma model, specifically 
\bea \label{Sp+1}
S^{(p+1)}&=&\int_{\S_{p+1}}\left(Z_i\w \dd X^{i}- A_i\w\dd Y^{i}+\Pi^{ij}(X)Z_i\w A_j  -\frac 12 \partial_k\Pi^{ij}(X)Y^{k}\w A_i\w A_j \,+\right. \nn \\[4pt]
&&\qquad \,\,\,\,\,\,\,\left. +\,\frac 1{(p+1)!}R^{i_1\dots i_{p+1}}(X)A_{i_1}\w \dots \w A_{i_{p+1}}\right)+\int_{\S_{p+2}}X^{\ast}H\,,
\eea 
where the last term is a Wess-Zumino one, obtained as the pull-back of the $(p+2)$-form $H$ on $M$, 
\be 
X^{\ast}H=\frac 1{(p+2)!}H_{i_1\dots i_{p+2}}(X)\,\dd X^{i_1}\w\dots \w\dd X^{i_{p+2}}
\ee 
and supported on an open $(p+2)$-brane $\S_{p+2}$ whose boundary is $\S_{p+1}$. 
The $(p+2)$-form $H$ is further assumed to be closed, $\dd H=0$, so that its variation drops to the boundary and its contribution to the field equations is only through the map X and not its extension that is necessary to define the higher-dimensional term in \eqref{Sp+1}. As usual, the quantum theory is well-defined provided 
that the homology class $[X(\S_{p+1})] \in H_{p+1}(M)$ vanishes and that $H$ defines an integer cohomology class \cite{Figueroa-OFarrill:2005vws}. 
 
Although the action functional \eqref{Sp+1} is written on a local patch of the target space $M$, it can be naturally defined globally once the relevant target space connections are introduced. Then the apparently non-tensorial coefficient $\partial_k\Pi^{ij}$ is completed to a tensor and the full theory can be written without using a local coordinate system on the target. Since this is not necessary for the purposes of the present paper, we refer to \cite{Chatzistavrakidis:2021nom} where a complete discussion of the covariant formulation appears. 

Provided that $(M,\Pi,R,H)$ is a twisted R-Poisson manifold of order $p+1$, it was shown in \cite{Chatzistavrakidis:2021nom} that the theory given by \eqref{Sp+1} is invariant under the following set of gauge transformations: 
\bea 
\label{gt1} \d X^{i}&=&\Pi^{ji}\epsilon_{j}\,,\\[4pt]
\label{gt2} \d A_i&=&\dd\epsilon_i+\partial_i\Pi^{jk}A_j\epsilon_k \,,\\[4pt]
\d Y^{i}&=&(-1)^{p-1}\dd\chi^{i}+\Pi^{ji}\, \psi_j -\partial_j\Pi^{ik}\left(\chi^{j}A_k+Y^{j}\epsilon_k\right)+\frac 1{(p-1)!}R^{iji_1\dots i_{p-1}}A_{i_1}\dots A_{i_{p-1}}\epsilon_j\,,\nn\\ \label{gt3}\\[4pt] 
\d Z_i&=&(-1)^{p}\dd\psi_i+\partial_i\Pi^{jk}\left(Z_j\epsilon_k+\psi_jA_k\right) -\partial_i\partial_j\Pi^{kl}\left(Y^jA_k\epsilon_l-\frac 12 \, A_kA_l\chi^{j}\right) \, + \nn\\ 
&&\qquad \qquad \quad + \, \frac {(-1)^p}{p!} \, \partial_iR^{ji_1\dots i_{p}}A_{i_1}\dots A_{i_{p}}\epsilon_j-\frac 1{(p+1)!}\Pi^{kj}H_{ijl_1\dots l_p}\Omega^{l_1\dots l_p}\epsilon_k \,, \label{gt4}
\eea 
where wedge products between differential forms are implicit. 
It is observed that there are three gauge parameters $(\epsilon_i,\chi^{i},\psi_i)$ of form degrees $(0,p-2,p-1)$ respectively. The gauge transformation of the scalar fields is controlled by the Poisson structure $\Pi$. Notably, the only appearance of the components of the $(p+2)$-form $H$ is in the ultimate term of the highest differential form field $Z_i$. They are combined with the world volume $p$-form $\Omega^{l_1\dots l_p}$ defined as 
\bea \label{Omega}
\Omega^{l_1\dots l_p}=\sum_{r=1}^{p+1}(-1)^{r}\prod_{s=1}^{r-1}\dd X^{l_{s}}\prod_{t=r}^{p}\Pi^{l_tm_t}A_{m_t}\,,
\eea 
which contains all possible combinations of $\dd X$ and $\Pi(A)$ that yield a $p$-form. This is essentially tailor-made to cancel the contribution of the Wess-Zumino term to the gauge variation of $S^{(p+1)}$. 

The field equations obtained from the action for the twisted R-Poisson sigma model read
\bea 
\label{eom1} F^{i}&:=&\dd X^{i}+\Pi^{ij}A_j=0\,,
\\[4pt]
\label{eom2} G_{i}&:=&\dd A_i+\frac 12 \partial_i\Pi^{jk}A_j\w A_k=0\,,
\\[4pt]
\label{eom3} {\cal F}^{i}&:=&\dd Y^{i}+(-1)^{p}\Pi^{ij}Z_j+\partial_k\Pi^{ij}A_j\w Y^{k}-\frac 1{p!}R^{ij_1\dots j_p}A_{j_1}\w\dots\w A_{j_{p}}=0\,,
\\[4pt]
{\cal G}_{i}&:=&(-1)^{p+1}\dd Z_i+\partial_i\Pi^{jk}\,Z_j\w A_k-\frac 12 \partial_i\partial_j\Pi^{kl}\,Y^{j}\w A_k\w A_l\, +\nn\\[4pt] 
&& \,+ \,  \frac 1{(p+1)!}\partial_iR^{j_1\dots j_{p+1}}A_{j_1}\w\dots\w A_{j_{p+1}}+\frac 1{(p+1)!}H_{ij_1\dots j_{p+1}}\dd X^{j_1}\w\dots\w \dd X^{j_{p+1}}=0\,. \nn\\ \label{eom4}
\eea 
Using the first field equation, i.e. the one of the highest form $Z_i$, the gauge transformation rule of $Z_i$ can be rewritten in an equivalent and more useful form as 
\bea 
\d Z_i&=&(-1)^{p}\dd\psi_i+\partial_i\Pi^{jk}\left(Z_j\epsilon_k+\psi_jA_k\right) -\partial_i\partial_j\Pi^{kl}\left(Y^jA_k\epsilon_l-\frac 12 \, A_kA_l\chi^{j}\right) \, + \nn\\ 
&+& \, \frac {1}{p!} \, f_i^{i_1\dots i_{p}j}A_{i_1}\dots A_{i_{p}}\epsilon_j \, - \nn\\[4pt]
&-&\frac 1{(p+1)!}\Pi^{kj}H_{ijl_1\dots l_p}\sum_{r=1}^{p}(-1)^{r+1}\binom{p+1}{r+1}\prod_{s=1}^{r} F^{l_{s}}\prod_{t=r+1}^{p}\Pi^{l_tm_t}A_{m_t}\epsilon_k \,,
\eea 
where we have defined 
\be \label{fdef}
f_{i}^{i_1\dots i_{p+1}}:=\partial_iR^{i_1\dots i_{p+1}}+H_{i}{}^{i_1\dots i_{p+1}}\,,
\ee 
and we introduced the short-hand notation of raising the indices of the components of the $(p+2)$-form $H$ via the 2-vector $\Pi$, specifically 
\be 
H_{i}{}^{i_1\dots i_{p+1}}=\Pi^{i_1l_1}\dots \Pi^{i_{p+1}l_{p+1}}H_{il_1\dots l_{p+1}}\,,
\ee 
and accordingly for other index structures.

We can now readily observe that the transformation of $Z_i$ contains the field strength of the scalar fields $X^{i}$, which is the field equation of $Z_i$. Remarkably, although for the 2D model ($p=1$) this transformation only contains $F^{i}$ linearly and without it appearing together with the field $A_i$, this ceases to be true in every other dimension higher than two. In contrast, for example in 3D one finds that{\footnote{In this paper, we use the subset symbol $\supset$ to mean that the right-hand side appears in the full expression of the left-hand side along with other terms that are not shown. It will mostly be used to provide heuristic explanations that clarify the often complicated structure of the quantities we compute. }} 
\be 
\d Z_i\supset \frac 1 {2} H_{il}{}^{mk}F^{l}A_{m}\epsilon_{k}+\frac 1 {3!}H_{ilm}{}^{k}F^{l}F^{m}\epsilon_{k}\,. 
\ee  
Thus both a product of $F^{i}$ with $A_i$ appears as well as a quadratic term in the field equation. Clearly the situation becomes even more non-linear in higher dimensions. This general feature of this class of theories is unusual and it reproduces itself in the closure of the gauge algebra and the square of the BRST operator that we will encounter in the next section. Although in gauge theories we are used to having gauge algebras that only close on-shell or BRST operators that are nilpotent only on-shell, we are not aware of particular examples where products of field equations appear. This should not be discouraging however, since the general statements of on-shell closure or on-shell nilpotency are still valid. Therefore one expects that these features can still be treated within the BV/BRST formalism and we show in the next sections that this is indeed the case. 

\section{BRST/BV formalism for R-Poisson sigma models}
\label{sec3}

\subsection{Ghosts and the BRST operator} 
\label{sec31} 

Having reviewed the classical action functional, the gauge transformations and the field equations of the theory, the next step would be to prepare it for quantization. 
Therefore, we are interested in determining the classical BV action, which will be the solution to the classical master equation. We recall that the BV extension is necessary when the theory is reducible as a constrained Hamiltonian system and when the gauge algebra closes only on-shell or the BRST operator is nilpotent only on-shell. Both these features are present in the class of theories we study, as we now describe in more detail. 

The first step toward quantization is to construct the classical basis of fields and ghosts. The ghosts correspond to the gauge parameters of the theory, promoted to fields of ghost number 1. To avoid introducing too much new notation, we denote the ghosts with the same letters as the gauge parameters. Thus the degree-1 ghosts are $(\epsilon_{i},\chi^{i},\psi_{i})$. However, the theory contains differential forms of form degree greater than 1 and therefore there will necessarily exist gauge transformations that are not independent. This means that the theory is highly reducible as a constrained Hamiltonian system and we must introduce additional ghosts for ghosts that take care of this redundancy. Indeed, the ghosts for the higher differential forms $Y^{i}$ and $Z_{i}$ are $\chi^{i}$ and $\psi_{i}$ of form degree $p-2$ and $p-1$ respectively. Being differential forms themselves means that we must include in the theory fields of ghost degree 2, say $\chi^{i}_{(1)}$ and $\psi_{i}^{(1)}$ of differential form degree $p-3$ and $p-2$. This process continues until we reach the top ghosts for ghost for each of the $\chi$ and $\psi$ series, which will be spacetime scalars. Thus we find that the classical basis contains the fields 
\be \label{basis}
(X^{i}, A_{i}, Y^{i}, Z_{i}, \epsilon_{i}, \chi^{i}_{(r)}, \psi_{i}^{(r)})\,,
\ee   
where the counter $r$ takes values from 0 to $p-2$ for the $\chi$-series $\chi^{i}_{(r)}$ and from 0 to $p-1$ for the $\psi$-series $\psi_{i}^{(r)}$.{\footnote{This discrepancy in the range is irrelevant; one could just state that the upper value is $p-1$ and the ghost $\chi^{i}_{(p-1)}$ does not exist since otherwise it would have negative form degree.}} 
Thus the classical basis contains a total of $2p+4$ fields of diverse ghost and form degree. At this stage, this is in accordance with the AKSZ construction; for example in the 3D case where $p=2$ we find 8 fields (4 ordinary fields, 3 ghosts and 1 ghost for ghost) as expected for the Courant sigma model \cite{Ikeda:2000yq,Ikeda:2002wh,Hofman:2002jz,Roytenberg:2006qz}. We collect these fields along with their ghost and differential form degrees in Table \ref{table1}.

	\begin{table}
	\begin{center}	\begin{tabular}{| c | c | c | c | c | c | c | c |}
			\hline 
			\multirow{3}{5.2em}{Field/Ghost} &&&&&&& \\ & $X^{i}$ & $A_i$ & $Y^i$ & $Z_{i}$ & $\epsilon_i$ & $\chi^{i}_{(r)}$ & $\psi_i^{(r)}$ \\ &&&&&&& \\ \hhline{|=|=|=|=|=|=|=|=|}
			\multirow{3}{6.0em}{Ghost degree} &&&&&&& \\ & $0$ & $0$ & $0$ & $0$ & $1$ & $r+1$ & $r+1$ \\ &&&&&&& \\\hline 
			\multirow{3}{5.5em}{Form degree} &&&&&&& \\  & $0$ & $1$ & $p-1$ & $p$ & $0$ & $p-2-r$ & $p-1-r$ \\ &&&&&&&
			\\\hline 
	\end{tabular}\end{center}\caption{The fields and ghosts of the twisted R-Poisson sigma model in $p+1$ dimensions. The range of $r$ is $r=0,\dots,p-1$ and we make the identifications $\chi^{i}_{(0)}\equiv \chi^{i}$, $\psi^{(0)}_i\equiv \psi_i$, so that we use a collective notation for the $p-1$ ghosts $\chi$ and the $p$ ghosts $\psi$. Obviously, $\chi_{(p-1)}^{i}=0$, since this ghost does not exist.  }\label{table1}\end{table} 

Since all fields of the theory are bi-graded, having one grading $\textrm{fd}(\cdot)$ as differential forms and one grading $\textrm{gh}(\cdot)$ as ghosts, we must choose a sign convention for their commutation. Herewith, for any two fields (including, later on, antifields) $\varphi_1$ and $\varphi_2$ we shall use the convention 
\be 
\varphi_1\varphi_2=(-1)^{\textrm{gh}(\varphi_1)\textrm{gh}(\varphi_2)+\textrm{fd}(\varphi_1)\textrm{fd}(\varphi_2)}\varphi_2\varphi_1\,.
\ee 

Next we define the BRST operator on the fields and ghosts, denoted as $s_0$ and raising the ghost degree by 1. Its action on the fields is simply the gauge transformation rule appearing in \eqref{gt1}-\eqref{gt4} with the gauge parameters replaced by the corresponding ghosts. Since we use the same notation for ghosts, we do not repeat these expressions here. The BRST operator should be nilpotent on-shell, 
\be 
s_0^2 \,(\cdot)  \overset{!}\approx 0\,,
\ee 
where $(\cdot)$ is a placeholder for any field or ghost and $\approx$ denotes that the field equations of the theory have been taken into account, or in other words that the square of the BRST operator is proportional to equations of motion. This requirement fixes the BRST transformation of the ghost fields. In particular, observing that 
\be 
s_0^2X^{i}= \Pi^{lk}\partial_k\Pi^{ji}\e_l\e_j+\Pi^{ji}s_{0}\e_{i}\,,
\ee 
and since $X^{i}$ is a scalar and {we can require that BRST transformations
of $X^i$ and $\epsilon_i$ do not contain field equations}, it is directly observed that due to $\Pi$ being a Poisson bivector the BRST transformation of the ghost $\e_i$ is completely fixed to be 
\be 
s_0\e_i=-\frac 12 \partial_i\Pi^{jk}\e_j\e_k\,.
\ee 
One may then check that $s_0^2\e_i=0$, as it should. Knowing the BRST transformation of $\e_i$ allows us to compute the square of the BRST operator on $A_i$ and find 
\be \label{s02A}
s_0^2A_i=-\frac 12 \partial_i\partial_j\Pi^{jk}F^{l}\e_j\e_k\,.
\ee 
We observe that it is proportional to the field equation for $Z_i$ and thus it vanishes only on-shell. This already dictates that the BV formalism must be used. Following this logic for the rest of the fields, leads to the BRST transformations of the ghost in the $\chi$ and $\psi$ series. They all follow the same pattern and therefore they can be presented collectively as 
\bea 
s_0\chi^{i}_{(r)}&=&\dd\chi^i_{(r+1)}+\partial_k\Pi^{ij}\left(A_j\chi^k_{(r+1)}-\epsilon_j\chi^k_{(r)}\right)-(-1)^{p+r}\Pi^{ij}\psi_j^{(r+1)}+\nn\\[4pt]
&& -\,\frac{\beta_{(r)}}{(r+2)!(p-r-2)!}R^{ij_1\ldots j_{r+2}k_1\ldots k_{p-r-2}}\epsilon_{j_1}\ldots\epsilon_{j_{r+2}} A_{k_1}\ldots A_{k_{p-r-2}}\,,\label{s0chi}\\[4pt]
s_0\psi_i^{(r)} &=& \dd\psi_i^{(r+1)}+\partial_i\Pi^{jk}\left(A_j\psi_k^{(r+1)}-\epsilon_j\psi_k^{(r)}\right)+\nn\\
&& +(-1)^{p+r}\partial_i\partial_l\Pi^{jk}\left(\frac{1}{2}\epsilon_j\epsilon_k\chi^l_{(r-1)}-\epsilon_j A_k\chi^l_{(r)}-\frac{1}{2}A_j A_k \chi^l_{(r+1)}\right)\,-\nn\\[4pt]
&&+\frac{(-1)^p\beta_{(r)}}{(r+2)!(p-r-1)!}f_i^{j_1\ldots j_{r+2}k_1\ldots k_{p-r-1}} \epsilon_{j_1}\ldots\epsilon_{j_{r+2}} A_{k_1}\ldots A_{k_{p-r-1}}+\nn\\[4pt]
&&+ \sum_{s=1}^{p-r-1}\frac{(-1)^{p(s+1)}\beta_{(r)}}{(s+1)!(r+2)!(p-r-s-1)!}\tensor{H}{_{il_1\ldots l_s}^{j_1\ldots j_{r+2}k_1\ldots k_{p-r-s-1}}}\times\nn\\[4pt]
&& \qquad\qquad\times\epsilon_{j_1}\ldots\epsilon_{j_{r+2}}A_{k_1}\ldots A_{k_{p-r-s-1}} F^{l_1}\ldots F^{l_s}\,, \label{s0psi}
\eea 
where 
\be 
\beta_{(r)}=(-1)^{p+r(r+1)/2}\,.
\ee
A useful remark is that the fields $Y^{i}$ and $Z_i$ may be seen as the ``$-1$'' elements in the $\chi$ and $\psi$ series, as is confirmed by inspection of the degrees in Table \ref{table1}. Indeed, if we identify 
\begin{equation}
\chi^i_{(-1)}:=(-1)^{p+1}Y^i \qquad\text{and}\qquad \psi_i^{(-1)}:=(-1)^pZ_i\,,\label{ids1}
\end{equation}
then the general formulas \eqref{s0chi} and \eqref{s0psi} are identical to the BRST transformations of $Y^{i}$ and $Z_{i}$ for $r=-1$, given in \eqref{gt3} and \eqref{gt4} with the gauge parameters replaced by the corresponding ghosts. This includes the term in $s_0Z_i$ containing field equations explicitly. Note that none of the ghosts in the $\chi$-series contains explicit equation of motion terms in their BRST transformation, whereas all ghosts in the $\psi$-series do, save the top one which is anyway a scalar.  
 The advantage of this identification is that once we compute the square of the BRST operator on the ghosts, the one for the fields $Y^{i}$ and $Z_i$ simply follows. A straightforward calculation leads to the results 
\bea 
s_0^2\chi^i_{(r)} &=& -\frac{\beta_{(r+1)}}{(r+3)!(p-r-4)!}R^{i l j_1\ldots j_{r+3}k_1\ldots k_{p-r-4}}\epsilon_{j_1}\ldots\epsilon_{j_{r+3}}A_{k_1}\ldots A_{k_{p-r-4}} G_l-\nn\\[4pt]
&&-\,\frac{(-1)^p\beta_{(r+1)}}{(r+3)!(p-r-3)!}\partial_l R^{i j_1 \ldots j_{r+3} k_1 \ldots k_{p-r-3}}\epsilon_{j_1}\ldots\epsilon_{j_{r+3}}A_{k_1}\ldots A_{k_{p-r-3}} F^l+\nn\\[4pt]
&&+\sum_{s=1}^{p-r-2}\frac{(-1)^{(p+1)s}\beta_{(r)}}{(s+1)!(r+3)!(p-r-s-2)!}\tensor{H}{_{l_1\ldots l_s}^{ij_1\ldots j_{r+3}k_1\ldots k_{p-r-s-2}}}\times\nn\\[4pt]
&&\qquad\qquad\times \epsilon_{j_1}\ldots \epsilon_{j_{r+3}} A_{k_1}\ldots A_{k_{p-r-s-2}}F^{l_1}\ldots F^{l_s}+\nn\\[4pt]
&&+\,\partial_k\partial_l\Pi^{ij}F^k\left(A_j\chi^l_{(r+2)}-\epsilon_j \chi^l_{(r+1)}\right)+\partial_k\Pi^{ij}\left(G_j\chi^k_{(r+2)}-\psi_j^{(r+2)} F^k\right)\,,\label{s02chir}
\eea
for the $\chi$-series of ghosts, and 
\bea
s_0^2\psi_i^{(r)} &=& \frac{(-1)^{p}\beta_{(r)}}{(r+3)!(p-r-3)!}\partial_i R^{lj_1\ldots j_{r+3}k_1\ldots k_{p-r-3}}\epsilon_{j_1}\ldots\epsilon_{j_{r+3}}A_{k_1}\ldots A_{k_{p-r-3}} G_l-\nn\\[4pt]
&& -\,\frac{\beta_{(r)}}{(r+3)!(p-r-2)!}\partial_l\partial_i R^{j_1\ldots j_{r+3}k_1\ldots k_{p-r-2}}\epsilon_{j_1}\ldots\epsilon_{j_{r+3}}A_{k_1}\ldots A_{k_{p-r-2}} F^l+\nn\\[4pt]
&&+\sum_{s=0}^{p-r-3}\frac{(-1)^{p(s+1)}\beta_{(r)}}{(s+2)!(r+3)!(p-r-s-3)!}\tensor{H}{_{il_1\ldots l_s}^{mj_1\ldots j_{r+3}k_1\ldots k_{p-r-s-3}}}\times\nn\\[4pt]
&&\qquad\qquad\times G_m\epsilon_{j_1}\ldots\epsilon_{j_{r+3}}A_{k_1}\ldots A_{k_{p-r-s-3}}F^{l_1}\ldots F^{l_s}+\nn\\[4pt]
&&+2\sum_{s=1}^{p-r-1}\frac{(-1)^{p(s+1)+s}\beta_{(r)}}{(s+1)!(r+3)!(p-r-s-1)!}\partial_{(i}\tensor{H}{_{l_1)l_2\ldots l_s}^{j_1\ldots j_{r+3}k_1\ldots k_{p-r-s-1}}}\times\nn\\[4pt]
&&\qquad\qquad\times \epsilon_{j_1}\ldots\epsilon_{j_{r+3}}A_{k_1}\ldots A_{k_{p-r-s-1}}F^{l_1}\ldots F^{l_s}+\nn\\[4pt]
&&+\,\partial_i\Pi^{jk} G_j \psi_k^{(r+2)}+\partial_i\partial_j\Pi^{kl}F^j\left(A_k\wedge\psi_l^{(r+2)}-\epsilon_k\psi_l^{(r+1)}\right)+\nn\\[4pt]
&&+\,(-1)^{p+r}\partial_i\partial_l\Pi^{jk}G_j\left(A_k\chi^l_{(r+2)}-\epsilon_k\chi^l_{(r+1)}\right)+\nn\\[4pt]
&&+\,(-1)^{p+r}\partial_i \partial_j\partial_m \Pi^{kl} F^j\left(\frac{1}{2}A_k A_l\chi^m_{(r+2)}+\epsilon_k A_l\chi^m_{(r+1)}-\frac{1}{2}\epsilon_k\epsilon_l \chi^m_{(r)}\right)\,, \label{s02psir}
\eea
for the $\psi$-series of ghosts. We observe that in both cases the field equations of $Y^{i}$ and $Z_{i}$ appear in all terms on the right-hand side. Moreover, according to the discussion above, the square of the BRST operator on $Y^{i}$ is found to be 
\bea
s_0^2Y^i &=& \frac{1}{2(p-3)!}R^{i l j_1 j_2 k_1\ldots k_{p-3}}\epsilon_{j_1}\epsilon_{j_2}A_{k_1}\ldots A_{k_{p-3}} G_l +(-1)^p\partial_k\Pi^{ij}\left(\psi_j^{(1)} F^k-G_j\chi^k_{(1)}\right)+\nn\\[4pt]
&&+\,\frac{(-1)^p}{2(p-2)!}\partial_l R^{i j_1 j_2 k_1 \ldots k_{p-2}}\epsilon_{j_1}\epsilon_{j_2}A_{k_1}\ldots A_{k_{p-2}} F^l+(-1)^p\partial_k\partial_l\Pi^{ij}F^k\left(\epsilon_j \chi^l-A_k\chi^l_{(1)}\right)-\nn\\[4pt]
&&-\sum_{s=1}^{p-1}\frac{(-1)^{(p+1)s}}{2(s+1)!(p-s-1)!}\tensor{H}{_{l_1\ldots l_s}^{ij_1 j_2 k_1\ldots k_{p-s-1}}}\epsilon_{j_1}\epsilon_{j_2}A_{k_1}\ldots A_{k_{p-s-1}}F^{l_1}\ldots F^{l_s}\,. \label{s0Y}
\eea 
For the corresponding expression of $Z_i$, which could alternatively be calculated directly from \eqref{gt4}, we find
\bea 
s_0^2 Z_i &=& \frac{(-1)^p}{2(p-2)!} \partial_i R^{lj_1 j_2 k_1\ldots k_{p-2}}\epsilon_{j_1}\epsilon_{j_2}A_{k_1}\ldots A_{k_{p-2}} G_l+(-1)^p\partial_i\Pi^{jk} G_j \psi_k^{(1)}-\nn\\[4pt]
&-&\frac{1}{2(p-1)!}\partial_l \partial_i R^{j_1 j_2 k_1\ldots k_{p-1}}\epsilon_{j_1}\epsilon_{j_2}A_{k_1}\ldots A_{k_{p-1}} F^l+(-1)^p\partial_j\partial_i\Pi^{kl}F^j\left(A_k\psi_l^{(1)}-\epsilon_k\psi_l\right)+\nn\\[4pt]
&+&\partial_i\partial_l\Pi^{jk}G_j\left(\epsilon_k \chi^l - A_k\chi^l_{(1)}\right)
+  \frac{(-1)^{p}}{2}\partial_i\partial_l\Pi^{jk}\e_j\e_k{\cal F}^{l} \, -\nn\\[4pt] 
&-&\partial_i\partial_j\partial_m\Pi^{kl} F^j\left(\frac{1}{2}A_k A_l\chi^m_{(1)}+\epsilon_k A_l\chi^m+\frac{(-1)^p}{2}\epsilon_k\epsilon_l Y^m\right)\, + \nn\\[4pt]
&+& \sum_{s=0}^{p-2}\frac{(-1)^{p(s+1)}}{2(s+2)!(p-s-2)!}\tensor{H}{_{il_1\ldots l_s}^{mj_1j_2k_1\ldots k_{p-s-2}}}G_m\epsilon_{j_1}\epsilon_{j_2} A_{k_1}\ldots A_{k_{p-s-2}}F^{l_1}\ldots F^{l_s}+\nn\\[4pt]
&+&\sum_{s=1}^p\frac{(-1)^{p(s+1)+s}}{(s+1)!(p-s)!}\partial_{(i}\tensor{H}{_{l_1)l_2\ldots l_s}^{j_1j_2k_1\ldots k_{p-s}}}\epsilon_{j_1}\epsilon_{j_2} A_{k_1}\ldots A_{k_{p-s}}F^{l_1}\ldots F^{l_s}\,.  \label{s0Z}
\eea
This completes the calculation of the square of the BRST operator on all fields. Since in most cases it does not vanish off-shell, the BV formalism is necessary to solve the classical master equation. 

Nevertheless, before proceeding with the BV formalism, it is worth listing the fields and ghosts whose BRST transformation is already nilpotent off-shell. First,  we  saw  that this is the case for $X^{i}$ and $\epsilon_i$. 
There exist, however, two more ghosts with this property. These are the top ghosts in each of the $\chi$ and $\psi$ series, namely $\chi^{i}_{(p-2)}$ and $\psi_{i}^{(p-1)}$, both being spacetime scalars.   
The general formulas yield 
\bea 
s_0\chi^{i}_{(p-2)}&=&-\partial_k\Pi^{ij}\e_j\chi^{k}_{(p-2)}-\Pi^{ij}\psi_{j}^{(p-1)}-\frac{\beta_{(p-2)}}{p!}R^{ij_1\dots j_p}\e_{j_1}\dots \e_{j_p}\,, \\[4pt] 
s_0\psi_i^{(p-1)}&=&-\partial_{i}\Pi^{jk}\e_j\psi_k^{(p-1)}-\frac 12\partial_i\partial_l\Pi^{jk}\e_j\e_k\chi^{l}_{(p-2)}+\frac{(-1)^p\beta_{(p-1)}}{(p+1)!}f_{i}^{j_1\dots j_{p+1}}\e_{j_1}\dots\e_{j_{p+1}}\,. \quad\quad \,\,\, 
\eea 
Either by direct computation or simply by inspection of the results \eqref{s02chir} and \eqref{s02psir}, we find 
\be 
s_0^2\chi^{i}_{(p-2)}=0=s_0^2\psi_i^{(p-1)}\,.
\ee  
We conclude that only $4$ of the $2p+4$ fields in \eqref{basis}, naturally being the four scalars, have nilpotent BRST operator acting on them. Therefore, for these fields there is no need to modify this operator, or in other words the BRST and the BV operator are identical for them. Thus we denote 
\be \label{bvisbrst}
s X^{i}:=s_0 X^{i}\,, \quad s\e_i:=s_0\e_i\,, \quad s\chi^{i}_{(p-2)}=s_0\chi^{i}_{(p-2)}\,, \quad s\psi_i^{(p-1)}:=s_0\psi_{i}^{(p-1)}\,,
\ee 
and $s^2$ vanishes on these fields. 

\subsection{Antifields and the untwisted BV operator}
\label{sec32} 

To pave the way towards determining the solution of the classical master equation of a twisted R-Poisson sigma model, we could follow one of two equivalent ways. The first step is common in either of the two and it amounts to enlarging the space of fields and ghosts by inclusion of the corresponding antifields and antighosts. For any field $\varphi$ we denote them as $\varphi_{+}$ (or $\varphi^{+}$, depending on the index position.) These are fields such that 
\bea 
\textrm{gh}(\varphi)+\textrm{gh}(\varphi_{+})&=&-1\,,
\\[4pt]
\textrm{fd}(\varphi)+\textrm{fd}(\varphi_+)&=&p+1\,.
\eea 
The full set of $2p+4$ antifields and antighosts with their degrees appears in Table \ref{table2}. In total the fields and antifields are $4(p+2)$ in number, a multiple of 4. This is to be expected, since without the Wess-Zumino term one could have used the AKSZ contruction with source space the graded manifold $T[1]\S$ and would have constructed 4 superfields containing the sum of all fields and antifields of total degree (the sum of ghost and form degrees) $0, 1, p-1, p$ respectively. In particular, the superfield $\mathbf{X}^{i}$ of total degree 0 would contain $(X^{i},Z_{+}^{i},\psi_{+}^{i}{}_{(r)})$, the total degree-1 superfield $\mathbf{A}_{i}$ would contain $(A_i,\epsilon_i,Y^{+}_{i},\chi^{+}_{i}{}^{(r)})$, the total degree-$(p-1)$ superfield $\mathbf{Y}^{i}$ would contain $(Y^{i},\chi^{i}_{(r)},A_{+}^{i},\e^{i}_{+})$ and the total degree-$p$ superfield $\mathbf{Z}_{i}$ would contain $(Z_i,\psi_i^{(r)},X^{+}_{i})$. The BV action would then be of the same form as the classical action but with superfields instead of fields. The Wess-Zumino term given by the pull-back of the 4-form $H$ is the sole reason that this would not be sufficient to determine the correct BV action.    

	\begin{table}
	\begin{center}	\begin{tabular}{| c | c | c | c | c | c | c | c |}
			\hline 
			\multirow{3}{4em}{Antifield} &&&&&&& \\ & $X^{+}_{i}$ & $A_{+}^i$ & $Y^{+}_i$ & $Z_{+}^{i}$ & $\epsilon_{+}^i$ & $\chi^{+}_{i}{}^{(r)}$ & $\psi_{+}^i{}_{(r)}$ \\ &&&&&&& \\ \hhline{|=|=|=|=|=|=|=|=|}
			\multirow{3}{6.0em}{Ghost degree} &&&&&&& \\ & $-1$ & $-1$ & $-1$ & $-1$ & $-2$ & $-r-2$ & $-r-2$ \\ &&&&&&& \\\hline 
			\multirow{3}{5.5em}{Form degree} &&&&&&& \\  & $p+1$ & $p$ & $2$ & $1$ & $p+1$ & $r+3$ & $r+2$ \\ &&&&&&&
			\\\hline 
	\end{tabular}\end{center}\caption{The antifields and antighosts of the twisted R-Poisson sigma model in $p+1$ dimensions. The range of $r$ is the same as for the corresponding fields.}\label{table2}\end{table} 

Next, one could use the BRST transformations found in the previous section to write down the extension of the classical action $S^{(p+1)}$ of \eqref{Sp+1} by all terms that contain one antifield and subsequently extend this action with all allowed terms with two and more antifields such that the classical master equation is satisfied. Alternatively, one could directly determine the BV operator $s$, i.e. the extension of the BRST operator $s_0$ that is nilpotent off-shell, using the fact that its action on the antifields produces equations of motion. Here we will work in this latter approach. 

Let us describe our approach in a heuristic way before presenting the details of the procedure. We have already found in Section \ref{sec31} that in all cases when the square of the BRST operator on the fields does not vanish, it is proportional to the field equations $F^{i}$, $G_{i}$ and ${\cal F}^{i}$, the latter appearing only in $s_0^2Z_i$. Therefore, we will certainly need the antifields from Table \ref{table2} whose transformation gives these field equations. These are 
$Z^{i}_{+}$, $Y_{i}^{+}$ and $A_{+}^{i}$, whose BV transformation will contain 
\bea 
s Z^{i}_{+} &\supset& (-1)^{p+1}\,F^{i}\,, \label{sZ+pre}\\[4pt] 
sY_{i}^{+}&\supset& \, G_{i}\,, \\[4pt]
sA_{+}^{i}&\supset& (-1)^p{\cal F}^{i}\,, \label{sA+pre}
\eea 
among other terms that we will determine. The goal then is to extend the BRST transformations by terms proportional to these antifields such that the square of the resulting operator vanishes. However, one should be careful with two more issues. The first issue is that once $Z_{+}^{i}$- and $Y^{+}_{i}$-dependent terms are included in the transformation of some field, the lower field which transforms as the derivative of the previous field will contain terms proportional to $\dd Z_{+}^{i}$ and $\dd Y^{+}_{i}$. This issue is ameliorated by noting that  
\bea 
s\psi^{i}_{+}&\supset& (-1)^p\,\dd Z^{i}_{+}\,,\\[4pt]
s\chi^{+}_{i}&\supset& (-1)^p\, \dd Y^{+}_{i}\,,
\eea  
and so on for the $\chi$ and $\psi$ series, since in general
\bea 
s \psi_{+}^{i}{}_{(r)} &\supset& \,\dd \psi_{+}^{i}{}_{(r-1)}\,, \\[4pt]
s \chi^{+}_{i}{}^{(r)} &\supset& -\,\dd \chi^{+}_{i}{}^{(r-1)}\,. 
\eea 
The second issue regards the appearance of explicit field equations in $s_0Z_i$ and in fact in all ghosts of the $\psi$-series. One may then ask whether any of the antifields will also contain explicit field equations in their BV transformation. The answer is necessarily yes and it will turn out to be very important in determining the correct BV action. Crucially, we will find that $sA^{i}_{+}$ contains $Z_{i}$ dependence and this will lead to a modification of its BV transformation by explicit $F^{i}$-dependent terms. Higher antifields will also get corrected accordingly, but it will become obvious that this will not be crucial for finding the BV action and can be determined a posteriori. This feature of higher ghosts and antifields having BV operator that contains field equations is one that does not exist in ordinary AKSZ constructions.

In summary, this heuristic discussion establishes the strategy for determining the BV operator on the fields. First we recall that the BV operator, denoted in general as $s_{\text{\tiny{BV}}}$ should satisfy the following three properties:
\begin{enumerate}
	\item[I.] When antifields are set to zero, it  reduces to the BRST operator $s_0$.  
	\item[II.] It is strictly nilpotent, $s_{\text{\tiny{BV}}}^2=0$, without using the field equations. 
	\item[III.] It is obtained from a BV action as $s_{\text{\tiny{BV}}}\cdot=(S_{\text{BV}},\cdot)$, with respect to the BV antibracket $(\cdot,\cdot)$ defined in Appendix \ref{appa}.
\end{enumerate} 
Note that there can (and will) exist operators $s$ other than $s_{\text{\tiny{BV}}}$ that satisfy the first two properties. It is the third property that establishes the right $s=s_{\text{\tiny{BV}}}$ that corresponds to a solution of the classical master equation. Our strategy then goes as follows. Consider the square of the BRST operator and add terms linear in the antifields to $s_0$, say $s_1$ such that the field equations cancel. Then compute the square of the modified BRST operator $s_0+s_1$, which will also be proportional to the field equations in general. Modify the operator $s_0+s_1$ by some antifield-dependent $s_2$ and repeat the procedure until the point that the modified operator is nilpotent off-shell. Then properties I and II are addressed. Property III is nearly automatic in the untwisted case but harder to satisfy in the twisted case. The strategy would be to add all possible additional $H$-dependent terms and solve a complicated set of consistency conditions. In the following, we will apply the above strategy in the untwisted case in arbitrary dimensions and we will also solve explicitly the twisted case in three dimensions, addressing the complicated property III. 

Let us now apply this procedure, starting with the simplest case of $A_i$, for which the square of the BRST operator is given in \eqref{s02A}. Using \eqref{sZ+pre}, we refine it to 
\be 
(s_0+s_1)A_i=\dd\e_i+\partial_{i}\Pi^{jk}A_j\e_{k}-\frac {(-1)^p}{2} \partial_i\partial_l\Pi^{jk}Z_{+}^{l}\e_j\e_k\,.
\ee  
The square of the modified operator can be easily calculated; requiring that it vanishes fixes completely the transformation of $Z_+^{i}$ too. Specifically, for 
\be \label{sZ+}
s Z_{+}^{i}=(-1)^{p+1}F^{i}+\partial_{j}\Pi^{ik}Z_+^{j}\e_{k}\,,
\ee 
we find that $(s_0+s_1)^2A_{i}=0$ identically. Therefore the BV operator on $A_i$ is 
\be \label{sA}
sA_i=\dd\e_i+\partial_{i}\Pi^{jk}A_j\e_{k}-\frac {(-1)^p}{2} \partial_i\partial_l\Pi^{jk}Z_{+}^{l}\e_j\e_k\,.
\ee
However, one should now cross check that $s^{2}Z_{+}^{i}=0$ too. This is a non-trivial consistency check, whose validity is easily established via an easy calculation, using also the modified transformation of $F^{i}$ which is directly computed to be 
\be 
s F^{i}=\partial_k\Pi^{ji}F^{k}\e_j-\frac {(-1)^p}{2} \Pi^{ij}\partial_j\partial_k\Pi^{lm}Z_{+}^{k}\e_l\e_m\,.
\ee 
In this way we have determined the BV operator on $A_i$ and $Z_{+}^{i}$. The fact that the procedure stopped quickly is only a feature of the low degree differential form $A_i$. For the rest of the fields reducibility kicks in and the procedure must be repeated multiple times. 
Fortunately, the common pattern of their BRST transformation allows us to perform this task once for each of the $\chi$ and $\psi$ series, before turning to the fields $Y^{i}$ and $Z_{i}$. 

For all ghosts $\chi^{i}_{(r)}$, we find the BV transformation 
\bea
s\chi^{i}_{(r)}&=&\dd\chi^{i}_{(r+1)}+\sum_{s=0}^{p-r-2}\frac {(-1)^p}{s!}\partial_k\partial_{l_1}\dots \partial_{l_s}\Pi^{ij}\sum_{s'=0}^{p-r-s-2}(-1)^{s'}{\cal O}^{l_1\dots l_s}(s,s')\mathfrak{X}^{k}_{j}(s,s')+ \nn\\[4pt]
&+&\sum_{s=0}^{p-r-2}\frac 1{s!}(-1)^{p+r+s-1}\partial_{l_1}\dots \partial_{l_{s}}\Pi^{ij}\sum_{s'=0}^{p-r-s-2}(-1)^{s'}{\cal O}^{l_1\dots l_s}(s,s')\psi_j^{(r+s+s'+1)} + \nn\\[4pt]
&-&\sum_{t=0}^{\lfloor\frac{p-r-2}{2}\rfloor}\sum_{s=0}^{p-r-2t-2}\sum_{s'=0}^{p-r-s-2t-2}\sum_{t'=0}^{p-r-s-s'-2t-2}\frac{(-1)^{(t+1)p+s'}}{s!t!}\frac{\beta_{(t-1)}\beta_{(a-2)}}{a!(p-a-t)!}\times \nn\\[4pt] &\times& \partial_{l_1}\dots\partial_{l_s}R^{ij_1\dots j_{a}k_1\dots k_{p-a}}{\cal O}^{l_1\dots l_s}(s,s')\widetilde{\cal O}_{k_1\dots k_{t}}(t,t') \e_{j_1}\dots \e_{j_a}A_{k_{t+1}}\dots A_{k_{p-a}}\,, \label{schir}
\eea  
where we denote $a:=r+s+s'+t+t'+2$ and we define the following operators,
\bea 
{\cal O}^{l_1\dots l_s}(s,s')&=&\sum_{\substack{m_i=-1 \\[2pt]  1\le i\le s-1}}^{s'-1}\left(\prod_{u=1}^{s-1}\psi_{+}^{l_{u}}{}_{(m_{u})}\right)\psi_{+}^{l_s}{}_{(s'-s-\sum_{i=1}^{s-1}m_i)}\,, \nn\\[4pt]
\widetilde{\cal O}_{k_1\dots k_t}(t,t')&=&\sum_{\substack{m_i=-1 \\[2pt] 1\le i\le t-1 }}^{t'-1}(-1)^{\sum_{q=0}^{\lfloor t/2\rfloor -1}(1+m_{t-1-2q})}\left(\prod_{u=1}^{t-1}\chi^{+}_{k_{u}}{}^{(m_{u})}\right)\chi^{+}_{k_t}{}_{(t'-t-\sum_{i=1}^{t-1}m_i)}\,, \nn\\[4pt]
\mathfrak{X}^{k}_{j}(s,s')&=&\sum_{u=0}^{p-r-s-s'-2}\widetilde{\cal O}_{j}(1,u-2)\chi^{k}_{(r+s+s'+u)}\,,
\eea 
with starting values 
\bea 
{\cal O}(0,s')&=&\d_{0,s'}\,, \\[4pt]
\widetilde{\cal O}(0,t')&=&\d_{0,t'}\,.
\eea 
We observe that the operators ${\cal O}$ and $\widetilde{\cal O}$ contain all products of antighosts of the $\psi_{+}$ and $\chi^{+}$ series and their fusion appears in the last term of the BV operator for the ghosts $\chi^{i}_{(r)}$. A few further remarks are in order. In these formulas, the antighosts $\chi^{+}_{i}{}^{(r)}$ have been extended to include the values $r=-1,-2,-3$, which by inspection of Tables \ref{table1} and \ref{table2} are identified with 
\be 
\chi^{+}_{i}{}^{(-1)}\equiv (-1)^{p-1}Y_i^{+}\,,\quad \chi^{+}_{i}{}^{(-2)}\equiv (-1)^p A_i\,,\quad \chi^{+}_{i}{}^{(-3)}\equiv (-1)^{p-1}\epsilon_i\,.
\ee  
There is nothing deep about these identifications, it is just one that uniformizes the presentation of the diverse expressions. In particular, it does not mean that $A_i$ and $\epsilon_i$ are antighosts, but only that they can be alternatively included in the antighost series for presentation purposes. 

Similarly, for all ghosts $\psi_{i}^{(r)}$ we find the BV transformation
\bea
s\psi_{i}^{(r)}&=&\dd\psi_{i}^{(r+1)}+\sum_{s=0}^{p-r-1}\frac {(-1)^p}{s!}\partial_i\partial_{l_1}\dots \partial_{l_s}\Pi^{jk}\sum_{s'=0}^{p-r-s-1}(-1)^{s'}{\cal O}^{l_1\dots l_s}(s,s')\widetilde{\mathfrak{X}}_{jk}(s,s')+ \nn\\[4pt]
&+&\frac {(-1)^{p+r}}{2}\sum_{s=0}^{p-r-1}\frac {(-1)^s}{s!}\partial_i\partial_l\partial_{l_1}\dots \partial_{l_{s}}\Pi^{jk}\sum_{s'=0}^{p-r-s-1}(-1)^{s+s'}{\cal O}^{l_1\dots l_s}(s,s')\times \nn\\[4pt]
&\times&\, \sum_{t=0}^{p-r-s-s'-1}\sum_{t'=0}^{t}(-1)^{t'}\widetilde{\cal O}_j(1,t'-2)\widetilde{\cal O}_k(1,t-t'-2)\chi^{l}_{(t+r+s+s'-1)} + \nn\\[4pt]
&+&\sum_{t=0}^{\lfloor\frac{p-r-1}{2}\rfloor}\sum_{s=0}^{p-r-2t-1}\sum_{s'=0}^{p-r-s-2t-1}\sum_{t'=0}^{p-r-s-s'-2t-1}\frac{(-1)^{pt+s'}}{s!t!}\frac{\beta_{(t-1)}\beta_{(a-2)}}{a!(p-a-t+1)!}\times \nn\\[4pt] &\times&\, \partial_{l_1}\dots\partial_{l_s}\partial_iR^{j_1\dots j_{a}k_1\dots k_{p-a+1}}{\cal O}^{l_1\dots l_s}(s,s')\widetilde{\cal O}_{k_1\dots k_{t}}(t,t') \e_{j_1}\dots \e_{j_a}A_{k_{t+1}}\dots A_{k_{p-a+1}}\,,\quad \quad \,\,\, \label{spsir}
\eea 
where the only new operator that appears is defined as 
\be 
\widetilde{\mathfrak{X}}_{jk}=\sum_{u=0}^{p-r-s-s'-1}\widetilde{\cal O}_{j}(1,u-2)\psi_{k}^{(r+s+s'+u)}\,.
\ee  
Once again, the fusion of the operators ${\cal O}$ and $\widetilde{\cal O}$ appears in the last term. What remains is to determine the BV operator acting on the fields $Y^{i}$ and $Z_{i}$. 
These are however just special values of the above general formulas by means of the identifications in \eqref{ids1}. The above universal formulas give the desired result of the operator that satisfies properties I, II and III. This can be alternatively found via the AKSZ construction since we have set $H=0$ to find these expressions and thus the QP structure is restored. Nevertheless, it is worth emphasizing that the BV operator found via AKSZ would at face value look much more complicated than \eqref{schir} and \eqref{spsir}.  
These formulas organise the different terms in a neat and simple way and they are valid in any dimension. 

Turning on $H$, the task of finding a closed expression for the BV operator with all $H$-dependent terms included becomes complicated. Nevertheless, the strategy we employed can still be applied, at least in a case by case fashion. In general, the requirement I. and the fact that we have determined the form of the BRST operator for all fields including the $H$-dependence already indicates that $s\psi_i^{(r)}$ is modified to 
\be 
s\psi_i^{(r)}|_{H=0} \mapsto s\psi_i^{(r)}|_{H=0}+ \D s\, \psi_i^{(r)}\,, 
\ee 
where the additional $H$- and $F$-dependent term $\D s\,\psi_i^{(r)}$, which vanishes in absence of $H$,  is given as 
\bea 
\D s\,\psi_i^{(r)}&=&\sum_{s=1}^{p-r-1}\frac{(-1)^{p(s+1)}\beta_{(r)}}{(s+1)!(r+2)!(p-r-s-1)!}\tensor{H}{_{il_1\ldots l_s}^{j_1\ldots j_{r+2}k_1\ldots k_{p-r-s-1}}}\times\nn\\[4pt]
&& \qquad\qquad\times\epsilon_{j_1}\ldots\epsilon_{j_{r+2}}A_{k_1}\ldots A_{k_{p-r-s-1}} F^{l_1}\ldots F^{l_s}+ \nn\\[4pt]  &+&\sum_{t=0}^{\lfloor\frac{p-r-1}{2}\rfloor}\sum_{s=0}^{p-r-2t-1}\sum_{s'=0}^{p-r-s-2t-1}\sum_{t'=0}^{p-r-s-s'-2t-1}\frac{(-1)^{pt+s'}}{s!t!}\frac{\beta_{(t-1)}\beta_{(a-2)}}{a!(p-a-t+1)!}\times \nn\\[4pt] &\times&\, \partial_{l_1}\dots\partial_{l_s}H_i{}^{j_1\dots j_{a}k_1\dots k_{p-a+1}}{\cal O}^{l_1\dots l_s}(s,s')\widetilde{\cal O}_{k_1\dots k_{t}}(t,t') \e_{j_1}\dots \e_{j_a}A_{k_{t+1}}\dots A_{k_{p-a+1}}\quad \quad \nn\\[4pt] &+&
\D_i(H)\,,\nn
\eea   
where the explicit contributions guarantee that property I. is satisfied and $\D_i(H)$ with $\D_i\overset{H\to 0}\longrightarrow 0$ has to be determined such that properties II. and III. are satisfied too. In addition, $s\chi^{i}_{(r)}$ is also modified with a corresponding term that should be determined. 
One should then apply the same algorithmic procedure of taking the square of the modified operator and refining it with suitable antifields as many times as necessary such that eventually its square vanishes. Once this is achieved, one must determine the relative weight of each of the unknown terms in the two series of $\chi$'s and $\psi$'s such that the nilpotent operator is indeed one obtained from a BV action through the antibracket. In the next section we apply this approach to the twisted R-Poisson sigma model in 3D.

\section{Twisted R-Poisson-Courant sigma models in 3D}
\label{sec4}

In this section we apply the general formalism developed above to a specific example, essentially the simplest non-trivial one that can be fully solved including the twist. This is a 3D Courant sigma model with a 4-form Wess-Zumino term. Such $H$-twisted Courant sigma models were considered from the viewpoint of first class constrained systems and 4-form-twisted Courant algebroids in \cite{Hansen:2009zd}. Here we study one such topological field theory that has the structure of a twisted R-Poisson sigma model. Apart from determining for the first time the BV action for twisted Courant sigma models, this task will be helpful in exemplifying (and of course extending to the twisted case) the rather complicated closed formulas derived in Section \ref{sec3}. 

 We consider the action functional \eqref{Sp+1} in three dimensions ($p=2$),  
 \bea \label{S3}
 S^{(3)}&=&\int_{\S_{3}}\left(Z_i\w \dd X^{i}- A_i\w\dd Y^{i}+\Pi^{ij}(X)Z_i\w A_j  -\frac 12 \partial_k\Pi^{ij}(X)Y^{k}\w A_i\w A_j \,+\right. \nn \\[4pt]
 &&\qquad \,\,\,\,\,\,\,\left. +\,\frac 1{3!}R^{ijk}(X)A_{i} \w A_{j}\w A_k\right)+\int_{\S_{4}}X^{\ast}H\,,
 \eea 
 with the Wess-Zumino term being the pull-back of a 4-form on the target space $M$, which is equipped with a twisted R-Poisson structure, consisting of a Poisson bivector $\Pi$ and an antisymmetric trivector $R$ that satisfy 
 \be 
 [\Pi,R]=\langle \otimes^4\Pi,H\rangle\,.
 \ee 
 In absence of $H$, this is a Bianchi identity for the derivation 
 \be 
 \dd_{\Pi} (\cdot):=[\Pi,(\cdot)]\,,
 \ee 
 which is nilpotent due to the Poisson condition $[\Pi,\Pi]=0$. In this case one notices that $Y^{i}$ is a spacetime 1-form and it may be combined with $A_i$ to a 1-form ${V}^{I}=(A_i,Y^{i})$ taking values in the pull-back of the generalized tangent bundle $TM\oplus T^{\ast}M$, where the index $I$ takes its $2\,\text{dim}\,M$ values. This observation is helpful in identifying the action \eqref{S3} with the general form of a Courant sigma model with Wess-Zumino term, which reads in our conventions as 
 \bea 
 S^{(\text{WZ-CSM})}&=&\int_{\S_{3}}\left(Z_i\w \dd X^{i}-\frac 12 \eta_{IJ}{V}^I\w\dd {V}^{J}+\rho^{i}_{I}(X)Z_i\w {V}^{I} \, + \right. \nn \\[4pt]
 &&\qquad \left.+\,\frac 1{3!} T_{IJK}(X) {V}^I\w {V}^J\w {V}^{K}\right)+\int_{\S_{4}}X^{\ast}H\,,
 \eea 
 with $\eta_{IJ}$ the $O(\text{dim}\,M,\text{dim}\,M)$ covariant metric
 \be 
 \eta=(\eta_{IJ})=\begin{pmatrix} 0 & \one_{\text{dim}\,M} \\ \one_{\text{dim}\,M} & 0\end{pmatrix}\,,
 \ee
  $\rho^{i}_{I}$ the components of the anchor map $\rho: E=TM\oplus T^{\ast}M\to TM$ of a Courant algebroid with vector bundle $E$ and $T_{IJK}$ the elements of the Courant bracket in a local basis. The example we use has anchor map components given by the Poisson bivector $\Pi$ and Courant bracket the twisted Koszul one. 
{For $H=0$, it is called a Poisson Courant algebroid or a contravariant Courant algebroid on a Poisson manifold \cite{Asakawa:2015jza, Bessho:2015tkk}.}
In presence of the Wess-Zumino term there is a departure from this Courant algebroid structure to a twisted one in the sense of \cite{Hansen:2009zd}, or a pre-Courant algebroid in the sense of \cite{Vaisman:2004msa}, which in our example becomes the twisted R-Poisson structure. More details on this relation are found in \cite{Chatzistavrakidis:2021nom}. 
 
 Our goal now is to determine the corresponding BV action of the classical action \eqref{S3}. According to our discussion in Section \ref{sec3}, there exist 16 fields and antifields, specifically the four fields $X^{i}, A_{i}, Y^{i}, Z_i$, their four antifields, three ghosts $\e_i, \chi^{i}, \psi_{i}$ and their three antighosts and one ghost for ghost $\widetilde{\psi}_i\equiv \psi_i^{(1)}$ and its antighost. First we briefly recall that when $H=0$ the BV action can be found using the AKSZ construction, see \cite{Roytenberg:2006qz}. In short, the above 16 fields are collected in four superfields of degrees $0,1,1,2$,
 \bea 
 \bs{X}^i&=& X^i +Z^{i}_{+} + \psi_{+}^{i}+\widetilde{\psi}_{+}^{i}~,\\ 
 \bs{A}_i&=&  \e_i+A_i+Y^{+}_{i}+\chi^{+}_{i}~, \\ 
 \bs{Y}^i&=&  \chi^{i}+Y^i +A_{+}^{i}+\e_{+}^{i}~,  \\
 \bs{Z}_i&=&  \widetilde{\psi}_{i}+\psi_{i}+Z_i+X^{+}_{i}~,
 \eea 
 defined on the graded Q-manifold $T[1]\S_{3}$ and taking values on the QP manifold $T^{\ast}[2]T^{\ast}[1]M$, which is isomorphic to $T^{\ast}[2]T[1]M$, which is typically associated to Courant sigma models. 
 Then the BV action is simply \cite{Roytenberg:2006qz}
  \bea 
  S^{(3)}_{\text{AKSZ}}=\int_{T[1]\S_{3}}\left(\bs{Z}_i \bs{\dd X}^{i}-\frac 12 \eta_{IJ}{\bs{V}}^I\bs{\dd} {\bs{V}}^{J}+\rho^{i}_{I}(\bs X)\bs Z_i {\bs V}^{I} +\frac 1{3!} T_{IJK}(\bs X) {\bs V}^I {\bs V}^J {\bs V}^{K}\right)\,,\,\,\,\,\, \label{Saksz}
  \eea 
 with $\bs V^{I}$ the superfield that combines $\bs A_i$ and $\bs Y^{i}$ and $\bs{\dd}$ the cohomological vector field on $T[1]\S_3$. 
 
 Once the twist $H$ is turned on though, this simple sequence of steps does not work, as already argued and as proven in \cite{Ikeda:2019czt} for the 2D AKSZ sigma model after twisting it by a 3-form. For the sake of completeness and for examplifying the general formulas of Section \ref{sec3}, we now present the BV operator on the eight fields as obtained by applying them in this case. First of all, due to \eqref{bvisbrst} and \eqref{sA}, we already have the BV operator on five of them, 
 \bea 
 sX^{i}&=&\Pi^{ji}\e_{j}\,, \\[4pt]
 s\e_i&=&-\frac 12 \partial_i\Pi^{jk}\e_j\e_k\,,\\[4pt]
 s\chi^{i}&=&-\partial_k\Pi^{ij}\e_j\chi^{k}-\Pi^{ij}\widetilde{\psi}_j-\frac 12R^{ijk}\e_j\e_k\,,\\[4pt]
 s\widetilde{\psi}_{i}&=&-\partial_i\Pi^{jk}\e_j\widetilde{\psi}_{k}-\frac 12 \partial_i\partial_l\Pi^{jk}\e_j\e_k\chi^{l}-\frac 1 {3!}f_i^{jkl}\e_{j}\e_k\e_l\,.\,\,\,\,\,\\[4pt]
 sA_i&=& \dd\e_i+\partial_{i}\Pi^{jk}A_j\e_{k}-\frac 12 \partial_i\partial_l\Pi^{jk}Z_{+}^{l}\e_j\e_k\,,
 \eea   
 where we recall that
 \be 
f_i^{jkl}= \partial_iR^{jkl}+\tensor{H}{_i^{jkl}}\,.
 \ee
We observe that in the above BV transformations only the one of the ghost for ghost $\widetilde{\psi}_i$ receives a correction due to the twist $H$, whereas the rest are identical to the AKSZ result. 
 For $Y^{i}$, partially guided by the formula \eqref{schir} for $p=2$ and $r=-1$ (recalling that $Y^{i}=-\chi^{i}_{(-1)}$), and adding a suitable $H$-dependent correction, we obtain 
 \bea 
 sY^{i}&=& -\dd \chi^{i}-\partial_k\Pi^{ij}(\e_jY^{k}+A_j\chi^{k})+\partial_k\partial_l\Pi^{ij}Z_{+}^{l}\e_j\chi^{k} + \nn\\[4pt]
 &&+\,\Pi^{ji}\psi_j+\partial_k\Pi^{ij}Z_{+}^{k}\widetilde{\psi}_j + \nn\\[4pt]
 &&+\,R^{ijk}\e_jA_k+\frac 12 \left(\partial_lR^{ijk}+\frac{1}{2}\tensor{H}{_l^{ijk}}\right)Z_{+}^{l}\e_j\e_k\,,
 \eea 
 where we wrote the terms exactly in order of appearance in \eqref{schir}. Note that the form of the $H$-correction in the ultimate term is absolutely necessary so as to satisfy all three required properties of the BV operator eventually. 
 Similarly, for $\psi_i$ we apply the formula \eqref{spsir} for $p=2$ and $r=0$  keeping the order of appearance and add suitable $H$-dependent terms to obtain
 \bea 
 s\psi_i&=&\dd\widetilde{\psi}_i+\partial_i\Pi^{jk}(-\e_j\psi_k+A_j\widetilde{\psi}_k)-\partial_i\partial_j\Pi^{jk}Z_{+}^l\e_j\widetilde{\psi}_k- \nn\\[4pt]
 &&-\, \partial_i\partial_l\Pi^{jk}(\e_jA_k\chi^{l}+\frac 12 \e_j\e_kY^{l})-\frac 12 \partial_i\partial_l\partial_m\Pi^{jk}Z_+^{m}\e_j\e_k\chi^{l} - \nn\\[4pt]
 &&+\frac{1}{4}\tensor{H}{_{il}^{jk}}\epsilon_j\epsilon_k F^l+\,\frac 12 f_{i}^{jkl}\e_j\e_kA_l-\frac 1{3!}\partial_{(m}f_{i)}^{jkl}Z_+^{m}\e_j\e_k\e_l\,.
 \eea 
 Finally, collecting together terms of the same type, for the field $Z_i$ we find  that  
 \bea 
 sZ_i&=&\dd\psi_i+\partial_i\Pi^{jk}(-\e_jZ_k+A_j\psi_k-Y^{+}_{j}\widetilde{\psi}_k) \,+ \nn\\[4pt]
 &+&\partial_i\partial_l\Pi^{jk}\left(\frac{1}{2}\epsilon_j\epsilon_k A_+^l -\epsilon_j A_k Y^l +\frac{1}{2}A_jA_k\chi^l+\epsilon_j \psi_k Z_+^l-A_j\widetilde{\psi}_k Z_+^l -\epsilon_j Y^+_k \chi^l +\epsilon_k \widetilde{\psi}_k \psi_+^l\right)+ \nn\\[4pt]
 &+&\partial_i\partial_l\partial_m\Pi^{jk}\left(\frac{1}{2}\epsilon_j\epsilon_k Y^l Z_+^m+\epsilon_j A_k\chi^l Z_+^m-\frac{1}{2}\epsilon_j \widetilde{\psi}_k Z_+^l Z_+^m +\frac{1}{2} \epsilon_j\epsilon_k\chi^l \psi_+^m\right)\, - \nn\\[4pt] 
 &-&\frac 14 \partial_i\partial_l\partial_m\partial_n\Pi^{jk}Z_+^mZ_+^n\e_j\e_k\chi^{l}+\frac{1}{2}f_i^{jkl}\epsilon_j A_kA_l+\frac{1}{6}\tensor{H}{_{ikl}^j}\epsilon_j F^k F^l+\frac{1}{2}\tensor{H}{_{il}^{jk}}\epsilon_j A_k F^l+\nn\\[4pt]
 &+&\partial_{(i}f_{m)}^{jkl}\left(\frac{1}{6}\epsilon_j\epsilon_k\epsilon_l\psi_+^m-\frac{1}{2}\epsilon_j\epsilon_k A_lZ_+^m\right)-\frac{1}{2}\left(\partial_i R^{jkl}+\frac{1}{2}\tensor{H}{_i^{jkl}}\right)\epsilon_j\epsilon_k Y^+_l- \nn\\[4pt]
&-&\frac{1}{6}\partial_{(i}\tensor{H}{_{m)l}^{jk}}\epsilon_j\epsilon_k F^lZ_+^m-\left(\frac{1}{12}\partial_{(m}\partial_n f_{i)}^{jkl}+\frac{1}{8}\partial_{(m}\partial_n\Pi^{jp}\tensor{H}{_{i)p}^{kl}}\right)\epsilon_j\epsilon_k\epsilon_l Z_+^mZ_+^n
 \eea 
 To verify that all the BV operators shown above are nilpotent off-shell, the complete ones for the antifields $Z_+^{i},Y^{+}_{i}, A_+^{i}$ and $\psi_{+}^{i}$ are needed too. 
 They are found to be 
 \bea 
 s Z_{+}^{i}&=&-F^{i}-\partial_{k}\Pi^{ij}\e_{j}Z_+^k\,, \\[4pt]
 sY^{+}_{i}&=&G_i-\partial_i\Pi^{jk}\e_jY^{+}_{k}+\partial_i\partial_l\Pi^{jk}\left(\frac 12 \e_j\e_k\psi_{+}^l-\epsilon_j A_k Z_+^l\right)-\frac 14 \partial_i\partial_l\partial_m\Pi^{jk}\e_j\e_kZ_+^{l}Z_{+}^{m}\,,\,\,\,\,\,\,\,\,\,\,\,\\[4pt]
 sA_+^{i}&=&{\cal F}^{i}-\partial_k\Pi^{ij}(\e_jA_+^{k}-Y^{+}_{j}\chi^{k}+\psi_jZ_+^{k}+\widetilde{\psi}_j\psi_+^k) -\,\nn\\[4pt]
 &&-\,\partial_k\partial_l\Pi^{ij}(A_j\chi^{k}Z_+^{l}+\e_jY^{k}Z_+^{l}-\frac 12 \widetilde{\psi}_{j}Z_+^kZ_+^l+\e_j\chi^k\psi_+^l) + \frac 12 \partial_k\partial_l\partial_m\Pi^{ij}\e_j\chi^kZ_+^lZ_+^m \,+\nn\\[4pt]
 && + \, R^{ijk}\e_jY^{+}_k+\left(\partial_l R^{ijk}+\frac{1}{2}\tensor{H}{_l^{ijk}}\right)\left(\epsilon_j A_k Z_+^l-\frac{1}{2}\epsilon_j\epsilon_k\psi_+^l\right)-\nn\\[4pt]
 &&-\frac{1}{6}\tensor{H}{_{kl}^{ij}}\epsilon_j F^k Z_+^l+\left(\frac{1}{4}\partial_{l}f_{m}^{ijk}-\frac{1}{12}\Pi^{in}\partial_l\tensor{H}{_{mn}^{jk}}\right)\epsilon_j\epsilon_k Z_+^l Z_+^m\\[4pt]
 s\psi_+^{i}&=&\dd Z_+^{i}+\Pi^{ij}Y_j^{+}+\partial_k\Pi^{ij}(A_jZ_+^{k}-\e_j\psi^{k}_{+})+\frac 12 \partial_k\partial_l\Pi^{ij}\e_jZ_+^kZ_+^l\,.
 \eea  
 Apart from confirming that the BV operator on the fields is nilpotent, a long yet straightforward calculation leads to the result that its action on these four antifields is also nilpotent, as desired. 
 
 With the above data, we can now write the candidate BV action for the 4-form-twisted R-Poisson sigma model in three dimensions. To present it in a compact way, let $\varphi^{\a}, \a=1,\dots,8$ be a collective notation for the eight distinct fields and ghosts of the theory, whose BV operator is given above. 
 The BV action is simply given as 
 \be \label{S3BV}
 S_{\text{BV}}^{(3)}=S^{(3)}-\sum_{\a}\int (-1)^{\text{gh}(\varphi)}\varphi^{+}_{\a}\,s_0\varphi^{\a}+\int \left(L_k\,Z_+^k+M_{kl}\,Z_+^{k}Z_+^{l}+N_{klm}\,Z_+^kZ_+^lZ_+^m\right)\,,
 \ee 
 with $S^{(3)}$ as in \eqref{S3} and 
 \bea 
 L_k&=& -\partial_k\Pi^{ij}\widetilde{\psi}_jY^{+}_i+\partial_k\partial_l\Pi^{ij}(\frac 12 \e_i\e_jA_+^l-\e_j\chi^kY^+_i+\e_i\widetilde{\psi}_{j}\psi_+^k)\,+\nn\\[4pt] 
 && +\, \frac 12 \partial_{k}\partial_l\partial_m\Pi^{ij}\e_i\e_j\chi^{l}\psi_+^m-\frac 12 (\partial_kR^{ijl}+\frac 12 H_{k}{}^{ijl})\e_j\e_lY^+_i+\frac 16 \partial_{(k}f_{m)}{}^{ijl}\e_i\e_j\e_l\psi_+^m\,,\,\,\,\,\,\,\,\,\, \\[4pt]
 M_{kl}&=& \frac 12 \partial_k\partial_l\Pi^{ij} (\e_i\psi_j-A_i\widetilde{\psi}_j)+\frac 12 \partial_k\partial_l\partial_m\Pi^{ij}(\e_iA_j\chi^m+\frac 12 \e_i\e_jY^m) - \nn\\[4pt]
 && - \,\frac 14 \partial_{(k}f_{l)}{}^{ijm}\e_i\e_jA_m-\frac 1{12}\partial_{(k}H_{l)m}{}^{ij}\e_i\e_jF^m\,,   \\[4pt]
 N_{klm}&=& -\frac 16 \partial_k\partial_l\partial_m\Pi^{ij}\e_i\widetilde{\psi}_j-\frac 1{12}\partial_{k}\partial_l\partial_m\partial_n\Pi^{ij}\e_i\e_j\chi^n- \nn\\[4pt] && -\, \left(\frac 1{36}\partial_{(k}\partial_lf_{m)}{}^{ijn}+\frac 1{24}\partial_{(k}\partial_l\Pi^{ip}H_{m)p}{}^{jn}\right)\e_i\e_j\e_n\,.
 \eea 
  That this is indeed the BV action, or in other words that it is the solution to the classical master equation $(S_{\text{BV}},S_{\text{BV}})$ with respect to the BV antibracket $(\cdot,\cdot)$ defined in Appendix \ref{appa}, can be seen as follows. The BV operator on the fields should satisfy the three properties I., II. and III. mentioned in section \ref{sec32}. 
To confirm that $S_{\text{BV}}$ as given in \eqref{S3BV} satisfies the classical master equation, it suffices to show that all the nilpotent operators $s$ derived above are indeed the unique BV operator stemming from $S_{\text{BV}}$ and moreover that the remaining four ones on the antifields of $X^{i},\e_{i},\chi^{i},\widetilde{\psi}_i$ are also strictly nilpotent. Then the classical master equation follows due to the graded Jacobi identity for the antibracket. This is not trivial because the operator $s$ can have additional off-shell ambiguities, terms that are proportional to the classical equations of motion of the theory. In particular, there are more than one ways to satisfy properties I. and II., and the point is to show that property III. completely fixes $s$ to be $s_{\text{\tiny{BV}}}$ without further ambiguities. 

 To show this, first of all notice that only terms proportional to the field equation $F^{i}$ constitute possible ambiguities. This is proven as follows. An ambiguity proportional to the field equation ${\cal G}_i$ can only potentially appear in the 3-form antifields $X^+_i, \e^{i}_+, \chi_i^{+}, \widetilde{\psi}_+^{i}$, since ${\cal G}_i$ is a 3-form; such ambiguity terms are a product of a scalar and ${\cal G}_i$. Since there are no scalar antifields, the ghost degree of a scalar that multiplies ${\cal G}_i$ has to be nonnegative which means that this type of correction can exist only for the antifields of ghost degree $-1$. The only such antifield is $X^+_i$ in which case the scalar multiplying ${\cal G}_i$ would have vanishing ghost degree, meaning that it is a function of $X$. But such terms in $sX^+_i$ are completely determined by the classical part of the BV action and cannot be modified which then eliminates all the ambiguities proportional to ${\cal G}_i$.

The ambiguities proportional to 2-form field equations $G_i$ and ${\cal F}^i$ are possible only in 3-form antifields $X^+_i$, $\epsilon_+^i$, $\chi^+_i$, $\widetilde{\psi}_+^i$, 2-form antifields $A_+^i$, $Y^+_i$, $\psi_+^i$ and a 2-form field $Z_i$. Here 2-form antifields cannot receive such corrections for the same reason why 3-form antifields could not receive corrections proportional to ${\cal G}_i$. In $sZ_i$ such corrections would not contain any antifields (because there are no scalar antifields), but all such terms are determined by the BRST operator (property I.). Finally, correction terms in 3-form antifields would have 1-form multiplying the field equation. This 1-form would need to contain an antifield, for otherwise, such a term would be determined by the classical part of the BV action. Since there are no scalar antifields and the only 1-form antifield is $Z_+^i$, the term would contain only $Z_+^i$ and no other antifields. However, this would produce terms in $sZ_i$ that contain no antifields and all such terms are determined by the BRST operator.

Finally, the only possibility are the ambiguities proportional to the field equation $F^i$. With this, 1-form fields $A_i$ and $Y^i$ cannot receive any corrections since those kind of terms cannot contain any antifields and as such are determined by the BRST operator. Similarly, $Z_+^i$ cannot receive those corrections as well because that would require $sZ_i$ to receive corrections that contain no antifields and that part is again determined by the BRST operator. On the other hand, there are no obstructions for $s\psi_i$ and $sZ_i$ to receive corrections proportional to $F^i$ (with the correction in $sZ_i$ containing at least one antifield). The remaining antifields would then receive corrections proportional to $F^i$ as well, but all those would be determined by the corrections of $s\psi_i$ and $sZ_i$. So, all the possible independent ambiguities are those proportional to the field equation $F^i$ in $s\psi_i$ and $sZ_i$. In addition, property I. is now completely satisfied. However, properties II. and III. still have to be taken into account. Taking into account all possible corrections, a straightforward calculation finally removes any remaining ambiguities.

 A final cross-check is to confirm that $s_{\text{\tiny{BV}}}^2$ vanishes on $X_i^{+},\e_{+}^{i},\chi_i^{+}$ and $\widetilde{\psi}^{i}_{+}$. First of all, using property III. we determine 
\bea 
s\widetilde{\psi}^{i}_{+}&=& \dd\psi_{+}^{i}+\Pi^{ji}\chi^{j}_+-\partial_k\Pi^{ij}(\e_j\widetilde{\psi}^k_{+}-Z_+^{k}Y^{+}_j-A_j\psi_+^k)\, + \nn\\[4pt] && + \,\partial_k\partial_l\Pi^{ij}(\e_jZ_+^l\psi_+^{k}-\frac 12 A_jZ_+^lZ_+^k)-\frac 1{3!}\partial_k\partial_l\partial_m\Pi^{ij}\e_jZ_+^kZ_+^lZ_+^m\,, \\[4pt]
s \chi_i^{+}&=& \dd Y_i^{+}+\partial_i\Pi^{jk}(-\e_j\chi^+_k+A_jY^+_k)+\, \nn\\[4pt] && +\, \partial_i\partial_l\Pi^{jk}(\e_jZ_+^lY^+_k+\e_jA_k\psi_+^l-\frac 12 \e_j\e_k\widetilde{\psi}_{+}^{l}+\frac 12 A_jA_kZ_+^l) +\, \nn\\[4pt]
&&+\,\frac 12 \partial_i\partial_l\partial_m\Pi^{jk}\e_j(\e_kZ_+^m\psi_+^l-A_kZ_+^lZ_+^m)-\frac 12 \partial_i\partial_l\partial_m\partial_n\Pi^{jk}\e_j\e_kZ_+^lZ_+^mZ_{+}^n\,, 
\eea
for the 3-form antighosts of the scalar ghosts of the theory, and moreover  
\bea 
s\e_i^{+}&=& -\dd A_+^{i}+\Pi^{ij}X^{+}_{j}-\partial_k\Pi^{ij}(\e_j\e_+^k-\chi^k\chi_j^{+}+\widetilde{\psi}_j\widetilde{\psi}_+^k+A_jA_+^k-Y^kY_j^+-\psi_j\psi_+^k+Z_jZ_+^k)\nn \\[4pt] 
&& -\, \partial_k\partial_l\Pi^{ij}(\e_j\chi^l\widetilde{\psi}_+^{k}+\e_jZ_+^lA_+^k+\chi^kZ_+^lY^+_j+\e_jY^l\psi_+^k+A_j\chi^l\psi_+^k-\widetilde{\psi}_{j}Z_+^l\psi_+^k+\nn\\[4pt] &&\qquad\qquad\, +A_jY^lZ_+^k-\frac 12 \psi_jZ_+^kZ_+^l)+ \nn\\[4pt] 
&& +\, \partial_k\partial_l\partial_m\Pi^{ij}(\e_j\chi^lZ_+^m\psi_+^k+\frac 12 \e_jY^lZ_+^kZ_+^m+\frac 12 A_j\chi^lZ_+^kZ_+^m-\frac 16 \widetilde{\psi}_{j}Z_+^kZ_+^lZ_+^m)- \nn\\[4pt] 
&&-\,\frac 16 \partial_k\partial_l\partial_m\partial_m\Pi^{ij}\e_j\chi^lZ_+^kZ_+^mZ_+^n-R^{ijk}(\e_j\chi_k^++A_kY^+_j)-\nn\\[4pt] 
&&-\,\partial_lR^{ijk}(\frac 12\e_j\e_k\widetilde{\psi}_+^l+\e_kZ_+^lY^+_j-\e_jA_k\psi_+^l-\frac 12 A_{j}A_kZ_+^l)+\\[4pt] && 
+\, \frac 12\partial_l\partial_mR^{ijk}( \e_j\e_kZ_+^l\psi_+^m-\e_jA_kZ_+^lZ_+^m)-\frac 1{12}\partial_l\partial_m\partial_nR^{ijk}\e_j\e_kZ_+^lZ_+^mZ_+^n+ \Delta s\,\e_i^{+}\,,\nn\\[4pt]
sX^{+}_i&=&(S_{\text{BV}},X_i^{+})\,,
\eea    
where we refrain from presenting the full result for $X^+_i$ since it contains all possible partial derivatives with respect to $X$ on every term of the BV action and is hence a very long expression. The $H$-dependent part of the transformation on $\e^+_i$ is hidden in $\D s$, which is given as
\bea 
\Delta s\,\e_+^i&=&-\frac 12 H_l{}^{ijk}(\e_j\e_k\widetilde{\psi}_{+}^{l}+\e_kZ_+^lY^+_j-2\e_jA_k\psi_+^l-A_jA_kZ_+^l)+\nn\\[4pt] 
&& +\, \frac 12 H_{kl}{}^{ij}F^{l}(\e_j\psi_+^k-A_jZ_+^k)-\frac 16 \partial_{(l}H_{m)k}{}^{ij}\e_jF^kZ_+^lZ_+^m+\frac 16 H_{jkl}{}^{i}F^kF^lZ_+^j+\nn\\[4pt] 
&& +\, \frac 12 \partial_{(m}H_{l)}{}^{ijk}(\e_j\e_kZ_+^m\psi_+^l-\e_jA_kZ_+^lZ_+^m)-\nn\\[4pt] &&-\,\left(\frac 1 {12}\partial_{(m}\partial_nH_{l)}{}^{ijk}+\frac 16 \partial_{(m}\partial_n\Pi^{ip}H_{l)p}{}^{jk}\right)\e_j\e_kZ_+^lZ_+^mZ_+^n\,.
\eea  Tracking all possible terms with an exterior derivative $\dd\cdot$ in the calculation of $s^2$ for any of these four antifields, we indeed find that they all vanish, as desired. 
 
 To facilitate the comparison with the BV operators and the BV action found through the AKSZ theory in the $H=0$ case, hence called $s_{\text{{\tiny{AKSZ}}}}$, we may rewrite the above expressions as 
 \bea 
 s\varphi^{\a}=s_{\text{{\tiny{AKSZ}}}}\varphi^{\a}+\Delta s\, \varphi^{\a}\,,
 \eea 
 where $\varphi^{\a}$ are the eight distinct fields and ghosts. Then $\Delta s$ vanishes for four of them, namely for $X^{i},\e_i,\chi^{i}$ and $A_i$, whereas for the remaining four we have found 
 \bea 
 \Delta s\,Y^{i}&=&\frac 14 H_{l}{}^{ijk}Z_{+}^{l}\e_j\e_k\,,\\
 \Delta s\, \widetilde{\psi}_i&=& -\frac 1 {3!}H_{i}{}^{jkl}\e_j\e_k\e_l\,, \\[4pt]
 \Delta s\, \psi_i&=&  \left(\frac 14 H_{il}{}^{jk} F^{l}+\frac 12 H_{i}{}^{jkl}A_l-\frac 1{3!}\partial_{(m}H_{i)}{}^{jkl}Z_{+}^m\e_l\right)\e_j\e_k \\[4pt]
 \Delta s\, Z_i&=&\left\{\frac 1{3!}H_{ikl}{}^{j}\, F^{k}F^{l}+\frac 12H_{il}{}^{jk}\, A_kF^l+\frac 14H_{i}{}^{jkl}\, (A_kA_l+\e_kY^{+}_l) +\frac{1}{3!}\partial_{(i}H_{m)l}{}^{jk} F^lZ_+^m\epsilon_k \right.  \,\quad  \nn\\[4pt] 
 && \left.+\frac 1{3!} \partial_{(m}H_{i)}{}^{jkl}\left(\epsilon_l\psi_+^m+3 A_lZ_+^m\right)\epsilon_k  
 -\frac 1{2\cdot 3!}\partial_{(m}\partial_n H_{i)}{}^{jkl}\epsilon_k\epsilon_l Z_+^mZ_+^n\right\}\e_j\,.
 \eea 
 This leads us to an alternative presentation of the BV action for the 4-form-twisted R-Poisson-Courant sigma model, which reads{\footnote{To avoid confusion, note that it is only $S_{\text{BV}}^{(3)}$ that satisfies the classical master equation. In the present context $S^{(3)}_{\text{{AKSZ}}}$ does not satisfy the classical master equation in general, but only when $H=0$. }} 
 \bea 
 S_{\text{BV}}^{(3)}&=&S_{\text{AKSZ}}^{(3)}+\Delta S^{(3)}\,,
 \eea 
 where $S_{\text{AKSZ}}^{(3)}$ is the AKSZ action for the untwisted R-Poisson-Courant sigma model given in \eqref{Saksz}, and $\Delta S^{(3)}$ is the $H$-dependent correction to it, given by 
 \bea 
 \Delta S^{(3)}&=& \int_{\Sigma_3}\left(-\frac{1}{6}\tensor{H}{_l^{ijk}}\epsilon_i\epsilon_j\epsilon_k\widetilde{\psi}_+^{l}-\frac{1}{4}\tensor{H}{_{kl}^{ij}}\epsilon_i\epsilon_j F^k\psi_+^l +\frac{1}{2}\tensor{H}{_l^{ijk}}\epsilon_i\epsilon_j A_k\psi_+^l+\frac{1}{6}\tensor{H}{_{jkl}^i}\epsilon_i F^jF^k Z_+^l\right. \nn\\[4pt]
 &+&\frac{1}{6}\partial_{(m}\tensor{H}{_{l)}^{ijk}}\epsilon_i\epsilon_j\epsilon_k Z_+^l\psi_+^m -\frac{1}{4}\tensor{H}{_l^{ijk}}\epsilon_i\epsilon_j Y^+_k Z_+^l+\frac{1}{2}\tensor{H}{_l^{ijk}}\epsilon_i A_jA_k Z_+^l- \nn\\[4pt]
 &-&\frac{1}{2}\tensor{H}{_{kl}^{ij}}\epsilon_i A_j F^k Z_+^l -\frac{1}{4}\partial_m\tensor{H}{_l^{ijk}}\epsilon_i\epsilon_j A_k Z_+^l Z_+^m+\frac{1}{12}\partial_m\tensor{H}{_{kl}^{ij}}\epsilon_i\epsilon_j F^k Z_+^l Z_+^m-\nn\\[4pt]
 &-&\left.\left(\frac{1}{36}\partial_m\partial_n\tensor{H}{_l^{ijk}}+\frac{1}{24}\partial_m\partial_n\Pi^{kp}\tensor{H}{_{lp}^{ij}}\right)\epsilon_i\epsilon_j\epsilon_k Z_+^l Z_+^m Z_+^n\right) +\int_{\S_4} X^{\ast}H\,.
 \eea 
 Obviously, when $H=0$ then $\D S^{(3)}=0$ and the correct solution of the classical master equation is given by the AKSZ action. 
 
\section{Conclusions}
\label{sec5}

The solution to the classical master equation of topological sigma models with a target space that possesses a QP structure as a graded manifold can be found using the AKSZ construction that provides at the same time a clear correspondence between geometry and field theory. In 2D this procedure results in the BV action of the Poisson sigma model and the A-/B-models and in 3D in the one of Chern-Simons theory and more generally of Courant sigma models. Higher-dimensional cases, essentially reflecting Hamiltonian mechanics in many dimensions, were formally discussed in \cite{Severa} and a 4D case was worked out completely in \cite{Ikeda:2010vz}.

In this paper, we studied topological sigma models whose target space does not have a genuine QP structure and therefore the systematic construction mentioned above does not apply as it is. This is motivated by the 2D example of the 3-form-twisted Poisson sigma model, where the Wess-Zumino term obstructs QP-ness of the target but the solution of the classical master equation was fully identified in \cite{Ikeda:2019czt}. Our main purpose was to generate new examples of this situation with an outlook towards developing a general geometric theory for the BV formalism of topological sigma models with Wess-Zumino terms. 
In this spirit, we started from the recently constructed twisted R-Poisson sigma models in arbitrary dimensions \cite{Chatzistavrakidis:2021nom}. In 3D this corresponds to 4-form twisted Courant sigma models \cite{Hansen:2009zd} (Chern-Simons theory with a Wess-Zumino term), whereas in general dimensions, say $p+1$, they correspond to twisting AKSZ models by a closed $(p+2)$-form. One of the advantages is that the theories are known in great detail and they offer the possibility of deriving explicit and universal formulas that are valid in any dimension, so they can be fully worked out.

Twisted R-Poisson sigma models are multiple stages reducible systems with an open gauge algebra. In a first step, we determined the BRST operator on all fields and ghosts of the theory in any dimension and by calculating its square we confirmed that it vanishes only on-shell. Notably, the square of the BRST operator is not linear in the classical field equations for all fields; instead, products of them can arise, a phenomenon that we could call ``non-linear openness of the gauge algebra''. To take care of this, we introduced the necessary antifields and antighosts dictated by the BV formalism. Our first main result then is that 
\bi 
\item in the untwisted case, namely when the Wess-Zumino term is turned off, we determined a complete set of expressions for the off-shell nilpotent BV operator of the theory that gives rise to the BV action that solves the classical master equation in any world volume dimension.  
\ei 
This result is formally not new, in the sense that these expressions could be derived using the AKSZ construction, since there is no obstruction to the QP structure on the target space in absence of Wess-Zumino term. Nonetheless, should one derive these formulas from the AKSZ/BV action, one would find an expanded and very complicated form of the expressions we derived. This is due to the fact that we followed a different strategy that may be summarized as follows. Instead of adding antifield-dependent terms in the classical action, which is the usual procedure in the BV formalism and in arbitrary dimensions it is a very hard thing to do, we instead followed a refinement procedure for the BRST operator. Specifically, knowing its square, we replaced each field equation in it with an antifield and added this term to the BRST operator. Calculating the square of the new operator, we find again terms proportional to the field equations. Repeating this procedure as many times as necessary, one can end up with an off-shell nilpotent operator, which becomes the BRST one once the antifields are turned off. Fortunately, due to repeating patterns in the transformation of the ghosts in the theory, this procedure is fully tractable. Requiring that the resulting operator is obtained from an action via the BV antibracket, we identify it with the BV operator of the theory. This procedure has the advantage that it yields elegant and closed expressions for the BV operator in any dimension in contrast to the AKSZ construction, while being equivalent to it. 

Once the Wess-Zumino term is turned on and as a result the R-Poisson structure is twisted, the procedure we suggested above requires to determine suitable modifications to the AKSZ/BV operator such that the new operator satisfies again all requirements to be a BV one, this time with the new geometric conditions brought about by the $(p+2)$-form twist. This is a hard problem, which we solve in 3D. Specifically, our second main result is that
\bi 
\item in the twisted case in three dimensions, we determined the full solution to the classical master equation. In other words we determined all necessary 4-form-dependent modifications to the BV operator and the BV action for 4-form-twisted R-Poisson-Courant sigma models.  
\ei 
This is then the second fully worked out example of a topological field theory with a non-QP target space whose BV action is identified, and the first in dimensions greater than two. We note that in two dimensions, there exist in fact many more examples based on Dirac sigma models, as reported in \cite{CJSS}. 

Based on the above results, it would be interesting to attempt the development of an extension to the AKSZ construction such that Wess-Zumino terms are taken into account and the target space geometry goes beyond QP structures. To achieve this, it would be helpful to solve the classical master equation in arbitrary dimensions and in presence of Wess-Zumino terms and also identify in every detail the higher geometric structures that appear in the problem, meaning all higher connections, torsion and curvature tensors that generalize the ones of the twisted Poisson sigma model. This would also be useful in solving the quantum master equation and identifying the corresponding quantum BV action for this class of theories, which would be also interesting in relation to deformation quantization. We plan to report on these issues in future work.

\appendix

\section{Antibracket conventions}
\label{appa}

An antibracket $(\cdot,\cdot)$ is defined as:
\begin{equation}
(F,G)=\int\dd^{p+1}\sigma\dd^{p+1}\sigma'\sum_\Phi\left(\frac{\delta_R F}{\delta\Phi(\sigma)}\frac{\delta_L G}{\delta\Phi^\ast(\sigma')}-\frac{\delta_R F}{\delta\Phi^\ast(\sigma)}\frac{\delta_L G}{\delta\Phi(\sigma')}\right)\delta(\sigma-\sigma')\,,
\end{equation}
where the sum goes over all fields and ghosts, and the antifields $\Phi^\ast$ are related to $\Phi^+$ through:
\begin{equation}
\Phi^+=\ast\Phi^\ast\,.
\end{equation} 
The right and left derivatives are defined as:
\begin{equation}
\delta S=\int\sum_\Phi \delta\Phi\frac{\delta_L S}{\delta\Phi}=\int\sum_{\Phi} \frac{\delta_R S}{\delta\Phi}\delta\Phi\,.
\end{equation}
From this follow useful identities:
\begin{eqnarray}
\left(\int a\Phi^+,\Phi\right) &=& (-1)^{p\cdot f(\Phi)}a\,,\\[4pt]
\left(\int b\Phi,\Phi^+\right) &=& -b\,,
\end{eqnarray}
where $f(\Phi)$ is the form degree of $\Phi$ and $a$ and $b$ have form and ghost degrees such that $a\Phi^+$ and $b\Phi$ are $(p+1)$-forms of vanishing ghost degree.

\paragraph{Acknowledgements.} We thank Larisa Jonke, Arash Ranjbar and Peter Schupp for useful discussions. Part of this work was done while one of us (A. Ch.) was visiting the Erwin Schr\"odinger International Institute for Mathematics and Physics in the framework of the Research in Teams Project: Higher Global Symmetries and Geometry in (non-)relativistic QFTs. Hospitality and financial support is gratefully acknowledged. 
This work is supported by the Croatian Science Foundation Project ``New Geometries for Gravity and
Spacetime" (IP-2018-01-7615) and by JSPS Grants-in-Aid for Scientific Research
Number 22K03323.


\begin{thebibliography}{99}
\bibitem{Batalin:1981jr}
I.~A.~Batalin and G.~A.~Vilkovisky,
``Gauge Algebra and Quantization,''
Phys. Lett. B \textbf{102} (1981), 27-31

\bibitem{Batalin:1983ggl}
I.~A.~Batalin and G.~A.~Vilkovisky,
``Quantization of Gauge Theories with Linearly Dependent Generators,''
Phys. Rev. D \textbf{28} (1983), 2567-2582
[erratum: Phys. Rev. D \textbf{30} (1984), 508]

	\bibitem{HT} 
M. Henneaux and C. Teitelboim,``Quantization of Gauge Systems'', Princeton University Press (1992).

\bibitem{Gomis:1994he}
J.~Gomis, J.~Paris and S.~Samuel,
``Antibracket, antifields and gauge theory quantization,''
Phys. Rept. \textbf{259} (1995), 1-145
[arXiv:hep-th/9412228 [hep-th]].

\bibitem{Witten:1988xj}
E.~Witten,
``Topological Sigma Models,''
Commun. Math. Phys. \textbf{118} (1988), 411
doi:10.1007/BF01466725

\bibitem{Witten:1991zz}
E.~Witten,
``Mirror manifolds and topological field theory,''
AMS/IP Stud. Adv. Math. \textbf{9} (1998), 121-160
[arXiv:hep-th/9112056 [hep-th]].

\bibitem{Deser:1981wh}
S.~Deser, R.~Jackiw and S.~Templeton,
``Topologically Massive Gauge Theories,''
Annals Phys. \textbf{140} (1982), 372-411
[erratum: Annals Phys. \textbf{185} (1988), 406]

\bibitem{Witten:1988hf}
E.~Witten,
``Quantum Field Theory and the Jones Polynomial,''
Commun. Math. Phys. \textbf{121} (1989), 351-399

\bibitem{Alexandrov:1995kv}
M.~Alexandrov, A.~Schwarz, O.~Zaboronsky and M.~Kontsevich,
``The Geometry of the master equation and topological quantum field theory,''
Int. J. Mod. Phys. A \textbf{12} (1997), 1405-1429
[arXiv:hep-th/9502010 [hep-th]].

\bibitem{SchallerStrobl}
P.~Schaller and T.~Strobl,
``Poisson structure induced (topological) field theories,''
Mod.\ Phys.\ Lett.\ A {\bf 9} (1994) 3129
[hep-th/9405110].

\bibitem{Ikeda}
N.~Ikeda,
``Two-dimensional gravity and nonlinear gauge theory,''
Annals Phys.\  {\bf 235} (1994) 435
[hep-th/9312059].

\bibitem{Ikeda:2000yq}
N.~Ikeda,
``A Deformation of three-dimensional BF theory,''
JHEP \textbf{11} (2000), 009
[arXiv:hep-th/0010096 [hep-th]].

\bibitem{Ikeda:2002wh}
N.~Ikeda,
``Chern-Simons gauge theory coupled with BF theory,''
Int. J. Mod. Phys. A \textbf{18} (2003), 2689-2702
\newline [arXiv:hep-th/0203043 [hep-th]].

\bibitem{Hofman:2002jz}
C.~Hofman and J.~S.~Park,
``BV quantization of topological open membranes,''
Commun. Math. Phys. \textbf{249} (2004), 249-271
\newline [arXiv:hep-th/0209214 [hep-th]].

\bibitem{Roytenberg:2006qz}
D.~Roytenberg,
``AKSZ-BV Formalism and Courant Algebroid-induced Topological Field Theories,''
Lett. Math. Phys. \textbf{79} (2007), 143-159
[arXiv:hep-th/0608150 [hep-th]].

\bibitem{Kontsevich:1997vb}
M.~Kontsevich,
``Deformation quantization of Poisson manifolds. 1.,''
Lett. Math. Phys. \textbf{66} (2003), 157-216
[arXiv:q-alg/9709040 [math.QA]].

\bibitem{Cattaneo:1999fm}
A.~S.~Cattaneo and G.~Felder,
``A Path integral approach to the Kontsevich quantization formula,''
Commun. Math. Phys. \textbf{212} (2000), 591-611
[arXiv:math/9902090 [math]].

\bibitem{Szabo:2001kg}
R.~J.~Szabo,
``Quantum field theory on noncommutative spaces,''
Phys. Rept. \textbf{378} (2003), 207-299
[arXiv:hep-th/0109162 [hep-th]].

\bibitem{Mylonas:2012pg}
D.~Mylonas, P.~Schupp and R.~J.~Szabo,
``Membrane Sigma-Models and Quantization of Non-Geometric Flux Backgrounds,''
JHEP \textbf{09} (2012), 012
[arXiv:1207.0926 [hep-th]].

\bibitem{Chatzistavrakidis:2015vka}
A.~Chatzistavrakidis, L.~Jonke and O.~Lechtenfeld,
``Sigma models for genuinely non-geometric backgrounds,''
JHEP \textbf{11} (2015), 182
[arXiv:1505.05457 [hep-th]].

\bibitem{Chatzistavrakidis:2018ztm}
A.~Chatzistavrakidis, L.~Jonke, F.~S.~Khoo and R.~J.~Szabo,
``Double Field Theory and Membrane Sigma-Models,''
JHEP \textbf{07} (2018), 015
[arXiv:1802.07003 [hep-th]].

\bibitem{Bessho:2015tkk}
T.~Bessho, M.~A.~Heller, N.~Ikeda and S.~Watamura,
``Topological Membranes, Current Algebras and H-flux - R-flux Duality based on Courant Algebroids,''
JHEP \textbf{04} (2016), 170
[arXiv:1511.03425 [hep-th]].

\bibitem{Heller:2016abk}
M.~A.~Heller, N.~Ikeda and S.~Watamura,
``Unified picture of non-geometric fluxes and T-duality in double field theory via graded symplectic manifolds,''
JHEP \textbf{02} (2017), 078
[arXiv:1611.08346 [hep-th]].

\bibitem{Severa:2016prq}
P.~\v{S}evera,
``Poisson-Lie T-duality as a boundary phenomenon of Chern-Simons theory,''
JHEP \textbf{05} (2016), 044
[arXiv:1602.05126 [hep-th]].



\bibitem{Klimcik:2001vg}
C.~Klimcik and T.~Strobl,
``WZW - Poisson manifolds,''
J. Geom. Phys. \textbf{43} (2002), 341-344
[arXiv:math/0104189 [math.SG]].

\bibitem{Kotov:2004wz}
A.~Kotov, P.~Schaller and T.~Strobl,
``Dirac sigma models,''
Commun. Math. Phys. \textbf{260} (2005), 455-480
[arXiv:hep-th/0411112 [hep-th]].

\bibitem{Chatzistavrakidis:2021nom}
A.~Chatzistavrakidis,
``Topological field theories induced by twisted R-Poisson structure in any dimension,''
JHEP \textbf{09} (2021), 045
[arXiv:2106.01067 [hep-th]].

\bibitem{Ikeda:2021rir}
N.~Ikeda,
``Higher Dimensional Lie Algebroid Sigma Model with WZ Term,''
Universe \textbf{7} (2021) no.10, 391
[arXiv:2109.02858 [hep-th]].

\bibitem{Alkalaev:2013hta}
K.~B.~Alkalaev and M.~Grigoriev,
``Frame-like Lagrangians and presymplectic AKSZ-type sigma models,''
Int. J. Mod. Phys. A \textbf{29} (2014) no.18, 1450103
[arXiv:1312.5296 [hep-th]].

\bibitem{Ikeda:2019czt}
N.~Ikeda and T.~Strobl,
``BV and BFV for the H-twisted Poisson sigma model,''
Annales Henri Poincare \textbf{22} (2021) no.4, 1267-1316
[arXiv:1912.13511 [hep-th]].

\bibitem{Blaom}
A.~D.~Blaom, \emph{Geometric structures as deformed infinitesimal symmetries}, Trans.~Am.~Math.~Soc.~{\bf 358} (2006) 3651.


\bibitem{Abad-Crainic} 
C.~A.~Abad and M.~Crainic, 
\emph{Representations up to homotopy of Lie algebroids},
Journal f\"{u}r die reine und angewandte Mathematik (Crelles Journal) {\bf 663} (2011) 91.


\bibitem{Kotov-Strobl2} 
  A.~Kotov and T.~Strobl,
  \emph{Lie algebroids, gauge theories, and compatible geometrical structures},
  Rev.\ Math.\ Phys.\  {\bf 31} (2018) no.04,  1950015.
  
  \bibitem{Hansen:2009zd}
  M.~Hansen and T.~Strobl,
  ``First Class Constrained Systems and Twisting of Courant Algebroids by a Closed 4-form,''
  Fundamental Interactions, pp. 115-144 (2009)
  doi:10.1142/9789814277839\_0008
  [arXiv:0904.0711 [hep-th]].
  
  \bibitem{Vaisman:2004msa}
  I.~Vaisman,
  ``Transitive Courant algebroids,''
  Int. J. Math. Math. Sci. \textbf{2005} (2005), 1737-1758
  [arXiv:math/0407399 [math.DG]].
  
  \bibitem{preca2}
  A.~J.~Bruce and J.~Grabowski,
  ``Pre-Courant algebroids,''
  arXiv:1608.01585 [math-ph].
  
  \bibitem{prehequiv}
  Z.-J.~Liu, Y.~Sheng and X.~Xu,
  ``The Pontryagin class for pre-Courant algebroids,''  
  J.\ Geom.\  Phys.  {\bf 104} (2016) 148--162
  [arXiv:1205.5898 [math-ph]].



\bibitem{Grewcoe:2020ren}
C.~J.~Grewcoe and L.~Jonke,
``Courant Sigma Model and $L_\infty$-algebras,''
Fortsch. Phys. \textbf{68} (2020) no.6, 2000021
[arXiv:2001.11745 [hep-th]].


\bibitem{Severa:2001qm}
P.~\v{S}evera and A.~Weinstein,
``Poisson geometry with a 3 form background,''
Prog. Theor. Phys. Suppl. \textbf{144} (2001), 145-154
[arXiv:math/0107133 [math.SG]].

\bibitem{Vaintrob}
A. Yu. Vaintrob, 
``Lie algebroids and homological vector fields,''
Russ. Math. Surv. 52 428 (1997)

\bibitem{Pham}
D.~N.~Pham, ``Higher Affine Connections,'' Mediterr. J. Math. 13, 12271262 (2016). https://doi.org/10.1007/s00009-015-0559-6

\bibitem{Kotov:2016lpx}
A.~Kotov and T.~Strobl,
``Lie algebroids, gauge theories, and compatible geometrical structures,''
Rev. Math. Phys. \textbf{31} (2018) no.04, 1950015
[arXiv:1603.04490 [math.DG]].

\bibitem{Figueroa-OFarrill:2005vws}
J.~M.~Figueroa-O'Farrill and N.~Mohammedi,
``Gauging the Wess-Zumino term of a sigma model with boundary,''
JHEP \textbf{08} (2005), 086
[arXiv:hep-th/0506049 [hep-th]].

\bibitem{Asakawa:2015jza}
T.~Asakawa, H.~Muraki and S.~Watamura,
``Gravity theory on Poisson manifold with $R$-flux,''
Fortsch. Phys. \textbf{63} (2015), 683-704
[arXiv:1508.05706 [hep-th]].


\bibitem{Severa} 
P.~\v{S}evera,
``Some title containing the words “homotopy” and
“symplectic”, e.g. this one,''
Travaux mathematiques, Volume 16 (2005), 121–137,
[arXiv:math/0105080].

\bibitem{Ikeda:2010vz}
N.~Ikeda and K.~Uchino,
``QP-Structures of Degree 3 and 4D Topological Field Theory,''
Commun. Math. Phys. \textbf{303} (2011), 317-330
[arXiv:1004.0601 [hep-th]].

\bibitem{CJSS}
A.~Chatzistavrakidis, L.~Jonke, T.~Strobl, G.~\v{S}imuni\'c, 
``Topological Dirac Sigma Models
and the Classical Master Equation,'' to appear.

\end{thebibliography}
\end{document}